\documentclass{article}
\usepackage[paperwidth=8.5in, paperheight=11in, margin=1in]{geometry}

\usepackage{datetime2}

\usepackage{algorithm}
\usepackage{algpseudocode}

\usepackage{graphicx}
\usepackage{caption}
\usepackage{subcaption}
\makeatletter
\newcommand*\bigcdot{\mathpalette\bigcdot@{.5}}
\newcommand*\bigcdot@[2]{\mathbin{\vcenter{\hbox{\scalebox{#2}{$\m@th#1\bullet$}}}}}
\makeatother

\usepackage{array} 
\usepackage{comment} 
\usepackage{afterpage}

\usepackage[geometry]{ifsym}

\usepackage{url}                       
\usepackage{hyperref}                 
\usepackage{bm}

\usepackage{amssymb, amsmath, amsfonts, latexsym}
\usepackage[mathscr]{euscript}

\usepackage{mathtools}

\usepackage{fancyref}

\usepackage{float}

\usepackage{rotating}

\usepackage{multirow}
\usepackage{array, makecell, rotating}

\usepackage{dirtytalk}

\usepackage{makecell} 

\def\indep{\perp\!\!\!\perp}

\font\Bigmath=cmsy10 scaled \magstep2

\def\diamondplustwo{\mathrel{%
  \ooalign{$+$\cr\hss\lower.255ex\hbox{\Bigmath\char5}\hss}}}

\usepackage{algpseudocode}

\newcommand{\calE}{\mathcal{E}}

\newcommand{\logit}{\text{logit}}

\usepackage{tikz}
\usetikzlibrary{arrows,decorations.pathmorphing,backgrounds,positioning,fit,petri}

\usepackage{tabu} 

\usepackage[square,sort,comma,numbers]{natbib}

\makeatletter \renewcommand\@biblabel[1]{#1.} \makeatother

\usepackage[pagewise]{lineno} 

\usepackage{fancyhdr}
\usepackage{lipsum}
\usepackage{background}
\backgroundsetup{
  position=current page.east,
  angle=-90,
  nodeanchor=east,
  vshift=-5mm,
  opacity=0.3,
  color=gray,
  scale=1.5,
  contents={Submitted Version (Stat Med DOI: 10.1002/sim.70290)}
}

\begin{document}

\begin{titlepage}
\begin{center}
  \bfseries
  \huge Model-Twin Randomization (MoTR) for Estimating the Recurring Individual Treatment Effect
  \vskip.2in
  \textsc{\large \href{https://www.ericjdaza.com/}{Eric J. Daza, DrPH, MPS} (\href{https://statsof1.org/}{Stats-of-1}), \href{https://evidation.com/}{Evidation})} \\
  \textsc{\large \href{https://igormatias.com/}{Igor Matias, MSc} (University of Geneva, \href{https://www.qualityoflifetechnologies.com/}{Quality of Life (QoL) Lab})} \\
  \textsc{\large Logan Schneider, MD (Stanford Medicine, Alphabet)} \\
  \textsc{\large \today}
  \vskip.2in
  \large {\sl Please send correspondence to:}
  \large Eric J. Daza (\href{mailto:ericjdaza@statsof1.org)}{ericjdaza@statsof1.org}) \\
\end{center}
%
%
%
%
\end{titlepage}
\clearpage

\section*{Abstract}

Temporally dense single-person \lq\lq small data" have become widely available thanks to mobile apps and wearable sensors. Many caregivers and self-trackers want to use these data to help a specific person change their behavior to achieve desired health outcomes. Ideally, this involves discerning possible causes from correlations using that person’s own observational time series data. In this paper, we estimate within-individual average treatment effects of physical activity on sleep duration, and vice-versa. We introduce the model twin randomization (MoTR; \lq\lq motor") method for analyzing an individual's intensive longitudinal data. Formally, MoTR is an application of the g-formula (i.e., standardization, back-door adjustment) under serial interference. It estimates stable recurring individual treatment effects, as is done in n-of-1 trials and single case experimental designs. We compare our approach to standard methods (with possible confounding) to show how to use causal inference to make better personalized recommendations for health behavior change, and analyze up to almost eight years of the authors' own Fitbit\texttrademark{} steps and sleep data.

\medskip
\noindent
{\bf Keywords:} causal inference, n-of-1, time series, longitudinal, digital, wearable, personalized
\clearpage

\section{Introduction}\label{sec:intro}

\say{I am large, I contain multitudes.} --- Walt Whitman

\medskip
The ever-increasing abundance of frequently collected, temporally dense single-person {\sl small data} \cite{2014-estrin, 2019-hekler-etal} has fueled the desire to extract \lq\lq personalized insights" from this digital individual-level information. However, standard analyses of such {\sl intensive longitudinal data} \cite{2006-walls-schafer} merely characterize statistical associations, making conclusions of causal effects generally hard to defend.

Causal effects are needed to make these insights truly \lq\lq actionable" to the extent that one actually expects to see an impact from intervening on a measured factor. Epidemiologists and econometricians have developed and used causal inference methods to better inform population-level policies and decisions to improve societal outcomes. How can clinicians and behavioral scientists likewise apply these methods to within-individual, single-person observational studies---thereby to find behavioral interventions and habit-changing practices that are most effective for a given individual?

One solution is to conduct within-individual experiments via mobile phone apps, sensors, and mail-in lab kits that enable small-data collection. These could be simple randomized or crossover designs as are commonly used in within-individual studies (see below), that do not generally require more nuanced causal inference methods.

Individuals with common recurring or chronic health conditions in particular could benefit from such digitally informed \lq\lq self-experiments". These conditions include migraines \cite{2019-schroeder-etal}, chronic pain \cite{2015_barr_etal}, irritable bowel syndrome \cite{2016-karkar-etal}, and many others \cite{2021-mcdonald-nikles, 2021-nikles-etal}. However, substantial barriers to self-experimentation exist for these conditions both methodologically and logistically (in the healthcare ecosystem) \cite{2022-selker-etal, 2023-selker-etal}. These barriers highlight the need to make better use of dense digital data by extracting insights that are not just \lq\lq correlational" (i.e., statistically associated) but also plausibly causal.


\subsection{Within-Individual Studies}

An n-of-1 trial is a randomized single-person crossover trial \cite{1994_matthews}; i.e., one individual undergoes multiple crossover periods with varying treatments \cite{2014_kravitz_etal, 2016_senn}. The target population in an n-of-1 trial is the largely unobserved superset of periods experienced by one person. Specifically, it is the set of all possible observation periods wherein the subject is at risk for the chronic health condition, and is under at least one of the treatment conditions being studied both during and outside of the n-of-1 trial total study period (i.e., before and after all n-of-1 treatment periods, in the \lq\lq real world"). Hence, we can speak of this target population as a \lq\lq population of yourself" or \lq\lq population of one" \cite{2018_daza}. An n-of-1 trial is known as a \lq\lq switchback experiment" in the business and finance literature \cite{2023-bojinov-etal}.

Biomedical research and clinical trials in particular have successfully employed n-of-1 trials \cite{1986_guyatt_etal, 1990_guyatt_etal, 1999_backman_harris, 2011_gabler_etal, 2011_lillie_etal, 2013_duan_etal}, with trial design, implementation, and analysis guidance available in various texts \cite{2014_naughton_johnston, 2014_kravitz_etal, 2015_nikles_mitchell, 2016_shamseer_etal, 2016_vohra_etal}. Chen et al (2012) \cite{2012_chen_etal} argued that wearable sensors can be used to facilitate n-of-1 trials. And both a recent article in Nature \cite{2015_schork} and the U.S. Department of Health and Human Services Agency for Healthcare Research and Quality (AHRQ) \cite{2014_kravitz_etal} have considered n-of-1 trials to be part of \lq\lq personalized medicine".

Single-case experimental designs (SCEDs) in psychology and n-of-1 trials in clinical settings are experimental single-person crossover studies focused on, though not always limited to, one individual. These two types of studies differ in design, but both are used to infer average individual-specific outcomes under different intervention levels. Each average is taken across a set of consecutive (though not necessarily contiguous) time intervals with the same exposure level. These intervals are called {\sl phases} in SCEDs, and {\sl periods} in n-of-1 trials.

Both SCEDs and n-of-1 trials share the same underlying statistical foundations and target estimands. In statistics, quantities that describe or apply to repeated measures taken on one study participant are described as \lq\lq within-subject" or \lq\lq within-individual". (Examples include the within-subject sum-of-squares, within-cluster variance, and the intra-cluster or intraclass correlation in mixed- or random-effects models and survey sampling.) Hence, we will collectively call SCEDs, n-of-1 trials, and their observational (i.e., non-experimental/non-randomized) counterparts \lq\lq within-individual" studies, similar to \lq\lq idiographic" (i.e., individualized/personalized) studies in psychology \cite{2005_ponterotto}.

Contrast these with hierarchical or multilevel models, which are used in group-level or \lq\lq nomothetic" studies to infer an average outcome taken across a set of repeatedly measured individuals. Their target population consists of a largely unobserved superset of individuals. For example, Araujo et al (2016) \cite{2016-araujo-etal} provide a good exposition on the relationship between n-of-1 trials and mixed-effects models. In the special case of a longitudinal study, longitudinal models are used to posit and infer a population-average trend taken across a set of individual trends. We elaborate on the relationship between within-individual studies and hierarchical models (and Bayesian inference) in the Appendix.

\subsection{Within-Individual Causal Inference}\label{subsec:within-individual-causal-inference}

We define a single-person crossover observational study to be a non-experimental, non-randomized study of one person, wherein recurring confounding and selection bias may exist \cite{2017-mcdonald-etal}. Our particular single-person crossover observational study objective is to discover plausible causal effects of would-be interventions that could be tested in a subsequent single-person crossover experiment like an n-of-1 trial or SCED.

Using the Neyman-Rubin-Holland counterfactual framework \cite{1923_neyman, 1974_rubin, 1986_holland} as an epistemological foundation, Daza (2018) \cite{2018_daza} formalized the single-person crossover study design in terms of statistical causal inference; specifically, to infer a single-person crossover average treatment effect (ATE) in either experimental (randomized) or observational studies. He called this within-individual ATE the {\sl average period treatment effect} (APTE), based on the n-of-1 trial \lq\lq period". The APTE can be thought of as a stable recurring individual treatment effect (RITE) \cite{2018_daza}.

The APTE framework is the first complete causal inference framework for n-of-1 studies. It elaborates on the implicit causal assumptions underlying the dynamic regression models of Vieira et al (2017) \cite{2017-vieira-etal}, specified for a single individual from the original formulation in Schmid (2001) \cite{2001_schmid} and Kravitz et al (2014) \cite{2014_kravitz_etal}. The framework enables a counterfactual interpretation of repeated effects over n-of-1 trial periods or SCED phases by stating common analytic challenges in terms of potential outcomes, including outcome autocorrelation and time trends (over multiple periods), and intervention carryover effects. It thereby gives statisticians, machine learning practitioners, data scientists, and regulators a solid conceptual foundation for building causal inference methods, deriving statistical properties, and defining causal estimands that include clinical endpoints and digital biomarkers.

Importantly, the APTE framework bridges and expands related areas of causal inference. It complements the causal inference theory of micro-randomized trials (MRTs) and just-in-time adaptive interventions (JITAIs), and enables counterfactual formalization of functional data analysis concepts \cite{2018_daza}. MRTs and JITAIs are close cousins to n-of-1 studies in that they also seek to personalize intervention effects. In the business and finance literature, Bojinov and Shephard \cite{2019-bojinov-shephard} formalized a potential-outcomes framework for randomized time series experiments. Recent work by Piccininni et al \cite{2024-piccininni-etal} has further expanded n-of-1 causal inference through their complementary \lq\lq U-CATE" framework.

Daza demonstrated how the APTE framework helped him infer putative causal effects using observational study data by applying two common causal inference methods, the g-formula (a.k.a., standardization, back-door or regression adjustment) and propensity-score inverse probability weighting, to infer a possible average effect of his physical activity on his weight. He used health activity data self-tracked over a six-year period. We elaborate on related approaches in Section \ref{subsec:related-approaches}.

In this paper, we propose a Monte Carlo method called model-twin randomization (MoTR, pronounced \lq\lq motor") for estimating the APTE. This approach involves first modeling the outcome as a function of all causes, and then running this \lq\lq model twin" through a simulated n-of-1 trial or experiment by randomizing the exposure at each time point to generate an effect estimate. MoTR is an application of the theory in Daza (2018) \cite{2018_daza} and the design-based approach of Wang (2021) \cite{2021_wang}. It has close theoretical connections to the innovative approach taken by Liang and Recht (2023) \cite{2023-liang-recht}, who use system identification from control theory to model the APTE as an impulse response. It is also closely related to microsimulation, as detailed in Section \ref{subsec:health-economics-microsimulation}.

\subsection{Why Study Sleep?}

The timing of sleep and wake is believed to be regulated by 2 main underlying processes: the homeostatic drive and the circadian rhythm \cite{1982-borbely}. The homeostatic drive is a use-state negative feedback process that accumulates over periods of wakefulness and dissipates over periods of sleep. The circadian rhythm is an intrinsic set of biological oscillations with approximately 24-hour (i.e., \lq\lq circa" means about; \lq\lq diem" means day) periodicity.

Each of these biological processes are regulated by not only the primary sleep period, but also by an organism's behaviors throughout the waking period. From the standpoint of the sleep homeostasis, the primary somnogen, adenosine, is a breakdown product of the energy molecule adenosine triphosphate (ATP), indicating that it is not just the amount of preceding wakefulness, but the intensity of that wakefulness (i.e., highly active vs sedentary) that dictates the subsequent sleep drive.

Similarly, there are a number of environmental exposures and biological processes (cumulatively termed zeitgebers or \lq\lq time givers") that help synchronize the body's intrinsic biorhythms to meet the organismal needs. Most of these signals---from dietary intake to light level and color variation---are integrated via a \lq\lq master clock" in the hypothalamus to orchestrate the molecular clocks present in every other cell of the body.

The aforementioned physiologic processes help to regulate the intrinsic sleep-wake state neurocircuitry in order to meet the needs of the organism (e.g., keeping nocturnal animals up at night). However, there is a wide range of variation in how these processes play out, even within species. Genome-wide association studies (GWAS) in humans have discovered polygenic contributions to the wide range of variation in typical sleep duration/need \cite{2021-garfield} and intrinsic circadian phase (i.e., \lq\lq chronotype" \cite{2017-kalmbach-etal}).

However, it seems that the intrinsic organismal needs also reflect the function that sleep serves in maintaining health. For example, families of individuals with mutations conferring short sleep need seem to sleep more effectively than others, as is evidenced by their physiologic and functional resilience to very short sleep durations \cite{2009-he-etal, 2019-shi-etal}. Studies of transgenic mice showing less neurodegeneration in response to chronic sleep restriction provide further evidence \cite{2022-dong-etal}. Taken together, there seems to be not only a genetically determined setpoint for sleep need and timing, but also an individual-specific resilience to perturbations of the sleep-wake system.

The current gold standards of sleep physiologic measurements are quite impractical: from the resource-intensive and disruptive nature of sleeping in a research laboratory to gather neurophysiologic, cardiopulmonary, and behavioral signals, to the highly controlled environment and inconvenient sampling requirements needed to isolate true circadian biomarker measurements (e.g., dim-light melatonin onset). On the other end of the spectrum, affording more real-world convenience at scale, self-report sleep and circadian measures---such as the Morningness-Eveningness questionnaire \cite{1976-horne-ostberg}, Munich Chronotype Questionnaire \cite{2003-roenneberg-etal}, and the Consensus Sleep Diary \cite{2012-carney-etal}---are plagued with recall bias and inaccuracy \cite{2019-mallinson-etal}. Toward this end, technologies that are able to gather sufficiently accurate data offer a unique opportunity for exploring the interrelationship between sleep and wake.

It is important to note that what is gained in convenience comes at the expense of what is lost in fidelity. For example, sleep-tracking technologies that depend on non-electroencephalographic signals don't accurately recapitulate in-lab polysomnograms \cite{2018-khosla-etal, 2021-chinoy-etal}. That being said, such sensors perform reasonably well in their general estimation of the daily amounts of sleep and wake \cite{2019-haghayegh-etal}, particularly in the generally healthy populations on which their algorithms were trained.

Similarly, different activity trackers have varying accuracy, depending upon context \cite{2020-montes-etal}. However, despite the known limitations in accuracy in how wearables measure both sleep and wake activities, their consistency in measurement longitudinally is core to the potential value they bring to understanding the impact of deliberate interventions or unintended changes within an individual.

For these reasons, examining sleep and how it relates to other behavioral metrics as measured by wearable sensors is an ideal target application for the development of idiographic causal inference methods. For example, one can examine varying the exposure to factors that impact the physiologic sleep drive to infer the APTE of that factor on sleep. Chevance et al (2021) \cite{2022-chevance-etal} examined such associations, but stopped short of attempting to draw causal inferences. A reasonable treatment phase or period could be one day, given that periods of sleep, by their very nature, tend to \lq\lq wash out" the effect of increased sleep pressure resulting from increased levels of waking activity. The target estimand of interest would be the true APTE of the exposure (e.g., activity level) on that particular night's sleep quantity; i.e., the average effect taken over all days throughout the individual's life.

\subsection{Paper Structure}

The rest of this paper is organized as follows. In Section \ref{sec:theory}, we define key notation and causal inference concepts. We introduce the autoregressive carryover model for estimating an APTE that is stable in the long run. We then introduce MoTR as a Monte Carlo (numerical) method for estimating the APTE in Section \ref{sec:methods}, and demonstrate its performance through a simulation study in Section \ref{sec:sims}. In Section \ref{sec:empirical}, we use MoTR (and a complementary propensity-score method) to estimate the APTE of walking cadence (i.e., steps per minute \cite{2018-tudor_locke-etal}) on sleep duration using up to almost eight years of EJD's and LS's Fitbit\texttrademark{} data. We conclude in Section \ref{sec:discuss} with a short summary, and propose future directions. All analyses were conducted in R version 3.6.3.

\section{Methodological Theory}\label{sec:theory}

\subsection{General Notation, Definitions, and Assumptions}\label{subsec:generalnotation}

Throughout this article, we define a {\sl data-generating function or mechanism} (or simply {\sl mechanism}) as the true, unknown equation that relates input variables or predictors to an outcome variable. We define a {\sl model} as a statistical equation that is fit to data, which may or may not approximate the true data-generating function. This distinction reflects the famous Box aphorism that \lq\lq all models are wrong, but some are useful". Our notation is as follows.

Random variables and fixed values are written in upper-case and lower-case, respectively. Let $p ( A = a )$ denote the probability mass or density of random variable $A$ at $a$, with shorthand $p ( a )$. For any random variable $B$, let $B | A $ denote the relationship \lq\lq $B$ conditional on $A$", with shorthand $B | a $ for $B | A = a$. Let $B \indep A$ denote statistical independence of $B$ and $A$. Let $E( \bigcdot )$ denote the general expectation operator, and let $I ( \bigcdot )$ denote the identity function such that $I(b) = 1$ if expression $b$ is true and $I(b) = 0$ otherwise. Let $E^{(m)}( \bigcdot )$ denote the empirical expectation operator over $t = 1, \hdots, m$ values, each with equal probability $\frac{1}{m}$.

We will use the following set notation. Let $\{ a \}$ denote a set with elements $\{ a_1, a_2, \hdots \}$. Let $\bm{a}$ denote a vector with elements denoted with commas as in $( a_1, a_2, \hdots )$, or without commas as in $( a_1 \, a_2 \, \cdots )$. Assume all vectors multiplied together are conformable; e.g., a $1 \times p$ random vector $\bm{A}$ and its $p \times 1$ coefficient vector $\bm{\beta}_A$ can be vector-multiplied as $\bm{A} \bm{\beta}_A$. Let $S^c$ represent the complement of set $S$.

Let $( \{ A_t \} ) = ( A_1, A_2, \hdots )$ denote a stochastic process or time series (i.e., vector of ordered random variables or vectors). Let $\overline{\bm{A}}_t^{\ell^A} = ( L^{\ell^A}, \hdots, L^2, L^1 ) A_t = ( A_{t-\ell^A}, \hdots, A_{t-2}, A_{t-1} )$ denote the history (i.e., set of all previous values of $A_t$) up to lag $\ell^A = 1, \hdots, t-1$, where $L^k$ is the lag operator $L^k A_t = A_{t-k}$. For example, $\overline{\bm{A}}_5^2 = ( L^2, L^1 ) A_5 = ( A_3, A_4 )$. To denote the full history from time point 1 up to and including time point $t-1$, we drop the $\ell^A$ notation and simply write $\overline{\bm{A}}_t = ( L^{t-1}, \hdots, L^1 ) A_t = ( A_1, \hdots, A_{t-1} )$. For example, $\overline{\bm{A}}_5 = ( L^4, L^3, L^2, L^1 ) A_5 = ( A_1, A_2, A_3, A_4 )$.

\subsection{Causal Inference}\label{subsec:causinf}

\subsubsection{Core Concepts}\label{subsec:causal-inference-core-concepts}

Let $Y$ represent a variable occurring after a binary variable $X$ with support $s = 0, 1$. If $E( Y | X = 0 ) \ne E( Y | X = 1 )$ when $X$ is randomized, then we say there is a {\sl direct effect} of $X$ on $Y$, and that $X$ is a {\sl direct cause} of $Y$. Let $\bm{W}$ represent the set of all other direct or indirect causes of $Y$ that may also be direct or indirect causes of $X$, or may {\sl contextualize} (i.e., modify or moderate) the effect of $X$ on $Y$.

An indirect cause affects the outcome through other mediating factors, forming a causal path from the indirect cause to the outcome through the {\sl mediators}. Causes of both $X$ and $Y$ {\sl confound} causal interpretations of the statistical dependence between $X$ and $Y$, and hence are called {\sl confounders}. An effect contextualizer is not directly manipulable, but the effect of $X$ on $Y$ varies across the contexts it represents (e.g., day of week).

The observable outcome $Y$ can then be written as a function of $X$, $\bm{W}$, and completely random error $\calE$ (e.g., due to random between-measurement variation). Let $Y = g ( X, \bm{W}, \calE )$ represent this {\sl structural causal mechanism} (i.e., true \lq\lq structural causal model" that generates the data) (SCM); for example, $g ( X, \bm{W}, \calE ) = \beta_0 + \beta_1 X + \bm{W} \bm{\beta}_2 + \calE$.

Now let $Y^s$ denote the value of $Y$ corresponding to $s$. This is the outcome that would be observed whenever $X = s$, and hence is called the {\sl potential outcome} (PO) of an exposure variable $X$. For $s = 0, 1$, the relevant potential outcomes are $Y^1$ and $Y^0$. The potential outcome $Y^s$ can then be written as a function of $\bm{W}$ and $\calE$. Let $Y^s = g_s ( \bm{W}, \calE )$ represent this PO mechanism. In design-based or randomization-based causal inference, the randomness in $Y$ is assumed to result only from the assignment of $X$ to 1 or 0; hence, the POs are treated as fixed values (implying $\bm{W}$ is also treated as fixed), and so are written simply as $y^1$ and $y^0$. We expand on why a potential outcome framework is useful in the Appendix.

The observed outcome $Y$, intervention of interest $X$, and potential outcomes $\big\{ Y^s \big\}$ are related by the equivalence $Y = \sum_{\{ s \}} Y^s I ( X = s )$, with shorthand $Y = Y^X$. This equivalence is called {\sl causal consistency}, and simply states that the observed outcome is equal to its corresponding PO. (We will refer to statistical consistency, the asymptotic unbiasedness of an estimator, simply as \lq\lq consistency".) Recalling our earlier example, we have $Y = Y^1 X + Y^0 (1 - X)$ because $s = 0, 1$. Under causal consistency, one natural set of PO mechanisms is $Y^1 = \beta_0 + \beta_1 + \bm{W} \bm{\beta}_2 + \calE$ and $Y^0 = \beta_0 + \bm{W} \bm{\beta}_2 + \calE$. The randomization-based counterpart to this model-based or superpopulation-based approach involves the same equations, but with fixed $\bm{w}$ and $\varepsilon$.

Causal consistency formalizes the {\sl fundamental problem of causal inference} \cite{1986_holland} that the only observable PO is the one corresponding to $X = s$. Hence, the term {\sl counterfactual outcome} (or simply {\sl counterfactual}) is often used to refer to $Y^{x'}$ for $x' \ne x$. \lq\lq Counterfactuals" and \lq\lq potential outcomes" are often used synonymously.

Statistical causal inference commonly involves two assumptions. (See the Appendix for a directed acyclic graph (DAG) \cite{1995_pearl, 2007_vanderweele_robins} that illustrates these assumptions.) When the POs occur independently of $X$, the assignment mechanism is said to be ignorable, and that {\sl ignorability} or unconfoundedness holds \cite{1983_rosenbaum_rubin, 1999b_greenland_etal}. We will see that this assumption is implied under randomization; i.e., $\big\{ Y^s \big\} \indep X$ if $X$ is randomized. If the POs occur independently of $X$ given all other causes $\bm{W}$, then {\sl conditional ignorability} holds; i.e., $\big\{ Y^s \big\} \indep X | \bm{W}$ \cite{2009_greenland_robins}.

{\sl Positivity} or {\sl overlap} is the requirement that all elements in the support of $X$ must necessarily co-occur with all other causes. For example, if $w = 0, 1$, then positivity holds if there is at least one observation for each unique pair of $\bm{w}$ and $x$. That is, positivity holds if each element of $\big\{ ( w, x ) \big\} = \big\{ (0, 0), (0, 1), (1, 0), (1, 1) \big\}$ is observed in the data. Positivity is required for estimating the average treatment effect, defined in Subsection \ref{subsec:ate}. Together, the ignorability, conditional ignorability, and positivity assumptions are often simply called \lq\lq ignorability" or {\sl strong ignorability} \cite{1983_rosenbaum_rubin}.

The second assumption that must hold is the stable unit treatment value assumption (SUTVA) \cite{1980_rubin}. A key component of this assumption is that the POs of a given individual are not affected by any other individual's treatment assignment. That is, there is no {\sl interference} between individuals \cite{1958_cox, 1990_rubin}.

Formally, let $\{ Y^1_i, Y^0_i \}$ and $\{ Y^1_{i'}, Y^0_{i'} \}$ denote the PO sets of distinct individuals $i$ and $i'$, respectively. If interference exists, then individual $i$ instead has POs $\{ Y^{11}_i, Y^{10}_i, Y^{01}_i, Y^{00}_i \}$, where $Y^{s_i s_{i'}}_i$ denotes the potential outcome if individual $i$ receives treatment $s_i$ while individual $i'$ receives treatment $s_{i'}$. Likewise holds for individual $i'$. In this case, causal consistency is defined as $Y = \sum_{\{ s_i, s_{i'} \}} Y^{s_i s_{i'}} I ( X_i = s_i, X_{i'} = s_{i'} )$, with shorthand $Y = Y^{X_i X_{i'}}$.

These studies require careful characterization, selection, and estimation of the relevant estimand \cite{2008_hudgens_halloran}. In Subsection \ref{subsec:apte}, we will see that autocorrelation and treatment carryover from past periods generate serial interference in single-person crossover studies. Wang (2021) \cite{2021_wang} and Liang and Recht (2023) \cite{2023-liang-recht} call the latter \lq\lq temporal interference".

A third assumption that is often overlooked but is important for establishing effect transportability (i.e., generalization of a treatment effect to a non-experimental setting) is called {\sl distributional stability} \cite{2002_dawid, 2005_dawid_didelez, 2010_dawid_didelez, 2012_eichler} or {\sl invariance} \cite{2020_buhlmann}. Invariance holds if the distribution of the outcome conditional on some subset of predictors does not change across all possible environments or intervention regimes (e.g., randomized vs. observed). Daza (2018) \cite{2018_daza} called the special case wherein randomization status neither affects the outcome nor depends on any confounders \lq\lq randomization invariance" (see Subsection \ref{subsec:g-formula}).

\subsubsection{Average Treatment Effect}\label{subsec:ate}

In a nomothetic study, participant $i$ with $\bm{W}_i = \bm{w}_i$ and $\calE_i = \varepsilon_i$ is thought to have a set of fixed POs $\{ y^s_i \}$ with the same cardinality as $\{ s \}$, as in design-based or randomization-based causal inference. That is, individual $i$ has the two fixed POs $y^1_i$ and $y^0_i$. A difference between $y^1_i$ and $y^0_i$ is called an {\sl individual treatment effect} (ITE).

The superpopulation-based counterparts to this randomization-based definition involve the same quantities, but with random $\bm{W}$ and $\calE$, yielding random POs $Y^1_i$ and $Y^0_i$. Here, the randomness in $Y$ is additionally assumed to result from other mechanisms such as random sampling of study participants, or random variation over time (even just throughout the study period). We will see in Subsection \ref{subsec:apte} onward that our time-series-based setting requires us to assume that at least some confounders are realizations of randomly varying quantities. Hence, we will expound on the superpopulation-based approach here to set up the theory for Subsection \ref{subsec:apte}.

While of primary interest, the ITE is not identifiable due to the fundamental problem of causal inference. However, if $X$ is randomized for all individuals, then $E( Y^s )$ (i.e., the mean PO taken across all individuals) is identifiable using only observed data because
\begin{align}
    E ( Y | X = x )
        &= E \left( Y^X | X = x \right) \text{ by causal consistency} \nonumber \\
        &= E \left( Y^x | X = x \right) \nonumber \\
        &= E \left( Y^x \right) \text{ when } X \text{ is randomized}
    	\label{eqn:porand}
.
\end{align}

A difference between $E( Y^1 )$ and $E( Y^0 )$ is called an {\sl average treatment effect} (ATE). Here, the expectations are taken over all study participants. In randomization-based causal inference, $E( Y^s )$ is taken across all study participants with unchanging confounders, contextualizers, and errors. In the superpopulation-based framework we will employ, $E( Y^s )$ is taken across the superpopulation consisting of all possible realizations of each study participant (i.e., the supports of $\bm{W}_i$ and $\calE_i$) across all participants. Contrast this with the more general superpopulation-based framework that treats the study participants as a sample of a larger target population.

Recall the example in Subsection \ref{subsec:causal-inference-core-concepts} with $Y^1_i = \beta_0 + \beta_1 + \bm{W}_i \bm{\beta}_2 + \calE_i$ and $Y^0_i = \beta_0 + \bm{W}_i \bm{\beta}_2 + \calE_i$. First, note that the ITE specified as $Y^1_i - Y^0_i$ is equal to $\beta_1$, regardless of individual. This is an important and common assumption in nomothetic causal inference when there is no interference: The ITEs are all equal to the same constant quantity, even though each potential outcome may be random. In general, we will assume that the ITE is a constant for each individual, as it is here. To indicate that this is a fixed quantity that is nonetheless comprised of random components, we will denote this using the lowercase $\delta^\text{ITE}_i = Y^1_i - Y^0_i$.

In this example, the ATE specified as $\delta^\text{ATE} = E( \delta^\text{ITE} ) = E( Y^1 - Y^0 ) = E( Y^1 ) - E( Y^0 )$ is also constant and equal to $\beta_1$. This expectation is taken over all individuals, who all have $\delta^\text{ITE}_i = \beta_1$, {\sl when $X$ is randomized for all individuals}. Hence, the ITEs are all equal to the ATE.

An ATE that varies based on some elements of $\bm{W}$ is called a conditional ATE (CATE) for heterogeneous treatment effects; i.e., the ATE is heterogenous across subgroups defined by values or levels of $\bm{W}$. The CATE is a more realistic estimand in many cases, and we will estimate an idiographic type of CATE (introduced in the next subsection) using real data in Section \ref{sec:empirical}.

The ATE is the primary estimand of interest in a randomized study like a randomized controlled trial (RCT) or A/B test because all other causes $\bm{W}$ need not be observed in order to estimate the difference between $E( Y^1 )$ and $E( Y^0 )$. That is, if $X$ is randomized for (i.e., randomly assigned to) each individual, then estimates of $E( Y | X = s )$---instead of $E( Y^s )$---can be used to estimate the ATE without needing to observe $\bm{W}$ or know how it functionally relates to $Y$. This result conveys the power of randomization as a tool for elucidating causal mechanisms. The key assumption, as shown in our example above, is that the ITEs are all identical.

In an observational study, it does not generally follow that $E \big( Y \big| X = x \big) = E \big( Y^x \big)$. This is because $\bm{W}$ may also affect $X$. When $X$ is not randomized, equation \eqref{eqn:porand} only holds true up to $E( Y | X = x ) = E( Y^x | X = x )$. Hence, estimates of $E \big( Y | X = 1 \big)$ and $E \big( Y | X = 0 \big)$ cannot be used to estimate the ATE. The effect of $\bm{W}$ on $X$ {\sl confounds} straightforward estimation of the ATE, so $\bm{W}$ is called a {\sl confounder}. If all confounders are observed, standard approaches to estimating the ATE from observational data can be used. We describe two of these in Section \ref{sec:methods}.

\subsection{Average Period Treatment Effect}\label{subsec:apte}

\subsubsection{Core Concepts}\label{subsec:apte-core-concepts}

In an idiographic study, a single participant $i$ is measured repeatedly over phases or periods $t = 1, \dots, m$. Henceforth, we will drop the $i$ index because we will only consider one participant. Let $Y_t$ represent the participant's recurring variable that occurs just after a preceding recurring binary variable $X_t$. Let $\bm{W}_t$ represent the set of all other causes of $Y_t$ that may also be causes of $X_t$. Let $Y_t = g ( X_t, \bm{W}_t, \calE_t )$ represent the structural causal mechanism.

The participant at period $t$ with $\bm{W}_t = \bm{w}_t$ and $\calE_t = \varepsilon_t$ is thought to have a set of fixed POs $y^1_t$ and $y^0_t$. Following Daza (2018) \cite{2018_daza}, we will call the ITE analogue in this setting a {\sl period treatment effect} (PTE), defined as a difference between $y^1_t$ and $y^0_t$. As mentioned earlier, we will develop our theory around the PTE superpopulation-based counterparts $Y^1_t$ and $Y^0_t$.

The PTE is not identifiable due to the fundamental problem of causal inference. However, if $X$ is randomized at every period, then $E( Y^s )$ (i.e., the mean PO taken across all periods) is identifiable using only observed data by the analogous derivation to \eqref{eqn:porand}. In this time series setting, we will invoke the sequential analogues of ignorability and conditional ignorability.

A difference between $E( Y^1 )$ and $E( Y^0 )$ is called an {\sl average period treatment effect} \cite{2018_daza}. In this paper, we will take this average over the superpopulation comprised of all $m$ observed periods. In the future, a broader goal that reflects those of nomothetic causal generalizability or transportability might be to estimate an APTE taken over all possible relevant periods throughout the participant's life (e.g., the times when they are at risk for condition $Y$ at varying levels of exposure $X$).

Let us now modify the example PO mechanisms in Subsection \ref{subsec:ate} by changing $i$ to $t$; i.e., $Y^1_t = \beta_0 + \beta_1 + \bm{W}_t \bm{\beta}_2 + \calE_t$ and $Y^0_t = \beta_0 + \bm{W}_t \bm{\beta}_2 + \calE_t$. As in our original example, the PTE specified as $\delta^\text{PTE}_t = Y^1_t - Y^0_t$ is equal to $\beta_1$ regardless of period. Analogous to the ATE, the APTE specified as
\begin{align}
    \delta^\text{APTE}
        &=
            E \left( \delta^\text{PTE} \right)
        \label{eqn:apte_simple}
\end{align}
is also constant and equal to $\beta_1$, where the expectation is taken over all periods {\sl when $X$ is randomized at every period}. More generally, we will assume that the PTE is constant over an entire period $t$, as implied by {\sl period-stable} predictor-outcome associations \cite{2018_daza}. We will not necessarily assume the PTE is constant across periods (as in our example), which is implied by {\sl stable} predictor-outcome associations \cite{2018_daza}.

This example reflects the common assumption in n-of-1 trials that the PTEs are all equal to the same constant quantity regardless of period. We will call this property {\sl effect constancy}; i.e., $\delta^\text{PTE}_t = \beta_1$ for all $t$ in our example. This follows because all associations between the predictors $X_t$ and $\bm{W}_t$ and the observed outcome $Y_t = \beta_0 + \beta_1 X_t + \bm{W}_t \bm{\beta}_2 + \calE_t$ are constant; i.e., predictor-outcome associations are stable. This property holds in n-of-1 trials with no serial interference across periods due to, for example, autocorrelation and carryover of the treatment's influence from past periods (see Subsection \ref{subsec:interference}).

The APTE is the primary estimand of interest in an n-of-1 trial because all other recurring causes $\bm{W}$ need not be observed in order to estimate the difference between $E( Y^1 )$ and $E( Y^0 )$. That is, if $X$ is randomized at each period (as in a standard n-of-1 trial), then estimates of $E( Y | X = s )$---instead of $E( Y^s )$---can be used to estimate the APTE without needing to observe $\bm{W}$ or know how it functionally relates to $Y$. The key assumption, as shown in our modified example, is that the PTEs are all identical when there is no serial interference.

However, we cannot generally ignore or mitigate autocorrelation and carryover in observational (i.e., non-experimental) real-world digital health settings. We will need to relax the assumption of effect constancy by allowing the PTE to vary across periods due to serial interference.

\subsubsection{N-of-1 Objective: CATE Expectations}\label{subsec:nof1objective}

In the causal inference literature, the CATE is a \lq\lq top-down" estimand of the ITE, wherein the two are generally equated \cite{2021-vegetabile}. It specifies ever-smaller groups of people as the number of predictors or covariates increase until, in the limit, only one person meets all conditions---at which point the fundamental problem of causal inference appears. In contrast (and complement), the APTE is a \lq\lq bottom-up" estimand that we believe more intuitively defines the ITE; one that starts with a single person measured repeatedly.

But there is no free lunch in causal inference. In the APTE framework, we are instead faced with a \lq\lq temporal fundamental problem" of causal inference at each time period, rather than for an entire individual. The framework allows us to estimate a stable average ITE---impossible to do with an RCT or A/B test---but not the ITE at a given time point.

Our overall research goal is to answer the behavior-change self-tracking question, \lq\lq What is a possible sustained effect (i.e., APTE) of $X$ on $Y$, that I might be able to modify?" For example, \lq\lq What is a possible average effect of walking quickly on my sleep duration if I keep this up over a week, versus walking slowly during that same week?" Here, we say an APTE is {\sl sustained} once it ceases to change after repeated assignment of the same treatment level over successive periods. This is the between-period equivalent of within-period {\sl effect stability} described in Daza (2018) \cite{2018_daza}.

Autocorrelation and carryover can delay an APTE from becoming sustained, and addressing these phenomena in order to answer our initial question is out of scope for this paper. Instead, we will first lay the theoretical groundwork by answering the easier question, \lq\lq What is the APTE of $X$ on $Y$ if we were to randomize $X$ at every period under similar conditions?" This can be answered by expanding the typical n-of-1 trial approach to allow autocorrelation and carryover to influence the APTE. Hence, we will describe such a study wherein $X$ is randomized at every period as an {\sl n-of-1 experiment} to distinguish it from the standard constraints enforced in an n-of-1 trial.

\subsection{Temporal Considerations}\label{subsec:tempconsids}

\subsubsection{Core Concepts}\label{subsec:temporal-considerations-core-concepts}

Relationships between variables across periods complicate estimation and interpretability of an APTE in ways not commonly encountered in estimating an ATE---even when interference is present. These complications arise from autocorrelation, time trends, carryover, and slow onset or decay \cite{2014_kravitz_etal}.

{\sl Autocorrelation} occurs if $Y_t$ depends on its history up to $\ell^Y$ lagged outcomes $\overline{\bm{Y}}_t^{\ell^Y}$. Note that $Y_t$ need not depend on all lags up to $\ell^Y$. For example, $Y_t$ could depend on $\overline{\bm{Y}}_t^4$ only through $( L^1, L^3, L^4 )$ or $( L^2, L^4 )$ for a given mechanism.

Redefine $\bm{W}_t = \bm{W}^{en}_t \cup \bm{W}^{ex}_t$ where $\bm{W}^{en}_t \cap \bm{W}^{ex}_t = \emptyset$. Here, $\bm{W}^{ex}_t$ denotes the set of exogenous causes of $Y_t$ such that $\bm{W}^{ex}_t$ never depends on either $X_t$ or $Y_t$. For example, a temporal cycle like weekday, week of month, month, annual quarter, or season could be included in $\bm{W}^{ex}_t$. Its complement is denoted $\bm{W}^{en}_t$, the set of endogenous causes of $Y_t$; i.e., $\bm{W}^{en}_t$ depends on either $X_t$ or $Y_t$. For example, $\overline{\bm{Y}}_t^{\ell^Y}$ would be included in $\bm{W}^{en}_t$ when autocorrelation is present.

We now write the general SCM of Subsection \ref{subsec:apte-core-concepts} as $Y_t = g \big( X_t, \bm{W}^{en}_t, \bm{W}^{ex}_t, \calE_t \big)$. Henceforth, we will focus on the simple case when the only possible endogenous causes of $Y_t$ are $\overline{\bm{Y}}_t^{\ell^Y}$ and $\overline{\bm{X}}_t^{\ell^X}$ such that $\bm{W}^{en}_t = \big\{ \overline{\bm{Y}}_t^{\ell^Y}, \overline{\bm{X}}_t^{\ell^X} \big\}$ and $Y_t = g \big( X_t, \overline{\bm{Y}}_t^{\ell^Y}, \overline{\bm{X}}_t^{\ell^X}, \bm{W}^{ex}_t, \calE_t \big)$.

For example, suppose $\ell^Y = 1$ such that $\overline{\bm{Y}}_t^{\ell^Y} = Y_{t-1}$, and suppose $\overline{\bm{X}}_t^{\ell^X} = \emptyset$. Hence, $Y_t = \beta_0 + \beta_X X_t + \beta_{ar} Y_{t-1} + \bm{W}^{ex}_t \bm{\beta}_{ex} + \calE_t$ where AR stands for \lq\lq autoregressive" (see Subsection \ref{subsec:arco}). For all such examples of linear models, we will assume $| \beta | < 1$ for all coefficients of lagged outcomes and their interactions with other variables. The latter is called the {\sl stationarity condition} in econometrics \cite{2000_hayashi}. We will also assume that $\big\{ ( \bm{W}^{ex}_t ) \big\}$ is jointly covariance stationary; equivalently, that joint weak- or wide-sense stationarity (WSS) of $\big\{ ( \bm{W}^{ex}_t ) \big\}$ holds. These two conditions imply that $\big\{ ( Y_t ) \big\}$ is WSS, ensuring our initial assumption is met.

A {\sl time trend} is defined as a sequential trend in the outcomes over successive periods such that $E( Y_t )$ increases or decreases across periods, rendering $\big\{ ( Y_t ) \big\}$ no longer WSS. For example, in $Y_t = \beta_0 + \beta_X X_t + \bm{W}^{ex}_t \bm{\beta}_{ex} + \beta_{tr} t + \calE_t$, the mean outcome $E( Y_t )$ increases by $\beta_{tr} > 0$ with every unit increase in $t$.

{\sl Carryover} is defined as \lq\lq the tendency for treatment effects to linger beyond the crossover (when one treatment is stopped and the next one started)" \cite{2014_kravitz_etal}. Suppose $Y_t$ depends on some set of lagged treatments $\overline{\bm{X}}_t^{\ell^X}$. If some lagged treatments in $\overline{\bm{X}}_t^{\ell^X}$ affect $Y_t$, then the SCM is $Y_t = g \big( X_t, \overline{\bm{X}}_t^{\ell^X}, \bm{W}^{ex}_t, \calE_t \big)$. In such cases, we will say that carryover or {\sl carryover influence} exists. In Subsection \ref{subsec:interference}, we will precisely define a carryover effect and see how it differs from carryover influence.

\subsubsection{Serial Interference and the Current Average Potential Outcome}\label{subsec:interference}

The presence of carryover or autocorrelation means that the POs at a given period are influenced by treatment exposure or assignment in past periods. That is, carryover can create {\sl serial interference} over time, as noted in the randomization-based temporal approaches of Wang (2021) \cite{2021_wang} and Aronow and Samii (2013) \cite{2013_aronow_samii}. Notably, we will see in the following example that carryover does not necessarily modify the PTE itself.

If carryover is present, recall that the SCM is $Y_t = g \big( X_t, \overline{\bm{X}}_t^{\ell^X},$ $\bm{W}^{ex}_t, \calE_t \big)$. To illustrate, suppose $\overline{\bm{X}}_t^{\ell^X} = X_{t-1}$ and $Y_t = \beta_0 + \beta_X X_t + \beta_{co} X_{t-1} + \beta_{Xco} X_t X_{t-1} + \bm{W}^{ex}_t + \calE_t$ where $\beta_{co}$ and $\beta_{Xco}$ collectively represents {\sl carryover influence}, and where $\beta_{co} = \beta_{Xco} = 0$ at $t=1$. Let $Y^{s_t s_{t-1}}$ denote the potential outcome if the participant receives treatment $s_t$ at period $t$, and treatment $s_{t-1}$ at period $t-1$. The PO mechanism is just $Y^{s_t s_{t-1}} = g_{s_t s_{t-1}} ( \bm{W}^{ex}_t, \calE_t ) = \beta_0 + \beta_X s_t + \beta_{co} s_{t-1} + \beta_{Xco} s_t s_{t-1} + \bm{W}^{ex}_t + \calE_t$. Serial interference exists, and the possible unique POs at period $t$ are $\{ Y^{11}_t, Y^{10}_t, Y^{01}_t, Y^{00}_t \}$. {\sl Serial causal consistency} states that $Y_t = Y^{X_t X_{t-1}}_t = \sum_{\{ s_t, s_{t-1} \}} Y^{s_t s_{t-1}}_t I ( X_t = s_t, X_{t-1} = s_{t-1} )$.

Let $\delta^\text{PTE}_t \big( s_{t-1} \big) = Y^{1 s_{t-1}}_t - Y^{0 s_{t-1}}_t$ denote the {\sl conditional PTE} defined as the PTE conditional on previous potential treatment level $s_{t-1}$, where we set $\delta^\text{PTE}_t \big( s_{t-1} \big) = Y^1_t - Y^0_t$ at $t=1$. Recalling the definition of \lq\lq carryover" from Subsection \ref{subsec:temporal-considerations-core-concepts}, it is reasonable to claim that a carryover effect exists if the current PTE differs based on past treatment level. We thereby define the {\sl carryover effect} of $X_{t-1}$ as $\delta^{co}_t = \delta^\text{PTE}_t (1) - \delta^\text{PTE}_t (0)$ at $t>1$; i.e., the difference between conditional PTEs.

To define the main PTE of interest, we will follow the approach Hudgens and Halloran (2008) \cite{2008_hudgens_halloran} used to define the \lq\lq individual average PO". Let $Y^{s_t \bigcdot}_t$ denote the {\sl current average PO} (CAPO) at period $t$ corresponding to $s_t$. In a future extension of this paper's methods involving multiple individuals, we might instead call this the {\sl contemporaneous average PO} across all individuals at a particular period.

The CAPO is the potential outcome at period $t$ under treatment level $s$ at that same period, averaged over all possible treatment level combinations over all past periods. In our example, the latter is just $s_{t-1} = 0, 1$. That is, $Y^{s_t \bigcdot}_t = E_{X_{t-1}} \big( Y^{s_t X_{t-1}} \big| X_t = s_t \big) = Y^{s_t 1}_t \Pr( X_{t-1} = 1 | X_t = s_t ) + Y^{s_t 0}_t \Pr( X_{t-1} = 0 | X_t = s_t )$. We thereby refine our definition of the PTE as $\delta^\text{PTE}_t = Y^{1 \bigcdot}_t - Y^{0 \bigcdot}_t$. Hudgens and Halloran (2008) \cite{2008_hudgens_halloran} might describe this quantity as a {\sl period average direct causal effect}.

\setlength{\extrarowheight}{4pt}
\begin{table}
\caption{ \label{tab:carrymod} Carryover example assuming $X = 0, 1$ is randomized with equal probability in an n-of-1 experiment. Carryover exists when either $\beta_{co} \ne 0$ or $\beta_{Xco} \ne 0$. Effect modification only exists when $\beta_{Xco} \ne 0$.}
\begin{center}
\scalebox{0.95}{
\begin{tabular}{ l | l | l }
\hline
Description & Term & Right-Side Expression \\
\hline
data-generating mechanism & $Y_t$ & $\beta_0 + \beta_X X_t + \beta_{co} X_{t-1} + \beta_{Xco} X_t X_{t-1} + \bm{W}^{ex}_t + \calE_t$ \\
PO & $Y^{11}_t$ & $\beta_0 + \beta_X + \beta_{co} + \beta_{Xco} + \bm{W}^{ex}_t + \calE_t$ \\
PO & $Y^{10}_t$ & $\beta_0 + \beta_X + \bm{W}^{ex}_t + \calE_t$ \\
PO & $Y^{01}_t$ & $\beta_0 + \beta_{co} + \bm{W}^{ex}_t + \calE_t$ \\
PO & $Y^{00}_t$ & $\beta_0 + \bm{W}^{ex}_t + \calE_t$ \\
conditional PTE & $\delta^\text{PTE}_t (1) = Y^{11}_t - Y^{01}_t$ & $\beta_X + \beta_{Xco}$ \\
conditional PTE & $\delta^\text{PTE}_t (0) = Y^{10}_t - Y^{00}_t$ & $\beta_X$ \\
carryover effect & $\delta^{co}_t = \delta^\text{PTE}_t (1) - \delta^\text{PTE}_t (0)$ & $\beta_{Xco}$ \\
CAPO & $Y^{1 \bigcdot}_t = Y^{11}_t 0.5 + Y^{10}_t 0.5$ & $\beta_0 + \beta_X + \beta_{co} 0.5 + \beta_{Xco} 0.5 + \bm{W}^{ex}_t + \calE_t$ \\
CAPO & $Y^{0 \bigcdot}_t = Y^{01}_t 0.5 + Y^{00}_t 0.5$ & $\beta_0 + \beta_{co} 0.5 + \bm{W}^{ex}_t + \calE_t$ \\
PTE & $\delta^\text{PTE}_t = Y^{1 \bigcdot}_t - Y^{0 \bigcdot}_t$ & $\beta_X + \beta_{Xco} 0.5$ \\
\hline
\end{tabular}
}
\end{center}
\end{table}
\setlength{\extrarowheight}{0pt}

Continuing our example, suppose $X$ is randomized to 0 or 1 with equal probability (i.e., all conditional probabilities are equal to 0.5), as in an n-of-1 experiment. Table \ref{tab:carrymod} lists the POs, conditional PTEs, carryover effect, CAPOs, and PTE. Suppose $\beta_{Xco} \ne 0$ at $t>1$. The carryover influence $\beta_{Xco}$ both produces a carryover effect (i.e., $\delta^{co}_t \ne 0$) and modifies the PTE (i.e., $\delta^\text{PTE}_t = \beta_X + \beta_{Xco} 0.5$).

Now suppose instead that $\beta_{Xco} = 0$ at $t>1$. There is no longer a carryover effect (i.e., $\delta^{co}_t = 0$), and the only possible non-zero carryover influence (i.e., $\beta_{co}$) does not modify the PTE. In this scenario, there is no effect modification---even though carryover exists! This highlights how, in general, a carryover influence implies neither a carryover effect nor effect modification.

\subsubsection{Other Temporal Considerations}\label{subsec:othertempconsids}

In an n-of-1 trial, carryover might be avoided in PTE estimation by using a {\sl washout period} that allows the lingering influence of past treatment to \lq\lq wash out" of the participant's system \cite{2014_kravitz_etal}. This is often done by design, through analytic adjustment, or both. The physical or design-based washout approach involves not administering the next treatment assignment until the influence of the previous treatment assignment has subsided. The analytic washout approach involves down-weighting or dropping time points at which carryover still exists.

In our example, an example of either approach involving dropping time points involves only including $t$ at which $X_{t-1} = 0$ (assuming $X = 0$ denotes the baseline treatment at which the outcome is at a baseline level). This constrains the PTE to just $Y^{10}_t - Y^{00}_t = \beta_X$ regardless of whether carryover modifies the overall PTE that allows for carryover. In this paper, we will assume we cannot designate a washout period a priori, and so must instead include any treatment history with potential carryover in our models.

If the participant is measured multiple times during a period or phase such that we have $j = 1, \dots, m_t$ repeated measurements per period $t$, there is an opportunity to assess and adjust for {\sl slow onset or decay} of the treatment effect. A treatment with slow onset takes time to reach its maximum or stable PTE, which may span more than one period. Similarly, a treatment with slow decay takes time to dissipate or wash out. If its washout time spans more than one period, then carryover exists. Multiple measurements per phase are standard practice in SCEDs. In studies that define a day as a period, intraday sensor measurements (as with a Fitbit\texttrademark{} or Apple Watch\texttrademark{}) are taken over uniform time intervals commonly called {\sl epochs}.

Slow onset and decay is a topic beyond the scope of this paper, and we will assume henceforth that there is no slow onset or decay. Still, in Section \ref{sec:discuss} we will propose a sketch of a longitudinal-based approach to model such intraday trends, with which to estimate an APTE that reflects a within-period trend.

\subsection{Deriving the APTE}\label{subsec:deriving-the-apte}

If there is no carryover, autocorrelation can still create serial interference. For example, suppose $\overline{\bm{Y}}_t^{\ell^Y} = Y_{t-1}$, $Y_1 = g \big( X_1, W^{ex}_1, \calE_1 \big)$, and $Y_t = g \big( X_t, Y_{t-1}, \bm{W}^{ex}_t, \calE_t \big)$ at $t > 1$. Even in this simple example, the resulting POs undergo a recursive combinatorial expansion with every successive period due to the autocorrelation. Estimating an APTE can easily become intractable with more and more periods.

To see this, note that we are faced with the sequence of PO mechanisms $Y^{s_1}_1 = g_{s_1} \big( \bm{W}^{ex}_1, \calE_1 \big)$, $Y^{s_2 s_1}_2 = g_{s_2 s_1} \big( Y^1_1 s_1 + Y^0_1 ( 1 - s_1 ), \bm{W}^{ex}_2, \calE_2 \big)$, and $Y^{s_3 s_2 s_1}_3 = g_{s_3 s_2} \big( Y^{1 s_1}_2 s_2 + Y^{0 s_1}_2 ( 1 - s_2 ), \bm{W}^{ex}_3, \calE_3 \big)$. For example, consider our earlier SCM for autcorrelation in Subsection \ref{subsec:temporal-considerations-core-concepts}, $Y_t = \beta_0 + \beta_X X_t + I(t>1) \big( \beta_{ar} Y_{t-1} \big) + \bm{W}^{ex}_t \bm{\beta}_{ex} + \calE_t$. Then $Y^{s_1}_1 = \beta_0 + \beta_X s_1 + \bm{W}^{ex}_1 \bm{\beta}_{ex} + \calE_1$, $Y^{s_2 s_1}_2 = \beta_0 + \beta_X s_2 + \beta_{ar} Y^{s_1}_1 + \bm{W}^{ex}_2 \bm{\beta}_{ex} + \calE_2$, $Y^{s_3 s_2 s_1}_3 = \beta_0 + \beta_X s_3 + \beta_{ar} Y^{s_2 s_1}_2 + \bm{W}^{ex}_3 \bm{\beta}_{ex} + \calE_3$, and so on.

\subsubsection{Historical Effects}\label{subsec:historical-effects}

Thankfully, the APTE can still be defined in the presence of serial interference. We will first introduce a type of APTE based only on the past treatment history. We will then see how to use the CAPO to derive the APTE as a function of this history-based quantity.

First, let us redefine the PO mechanism of Subsection \ref{subsec:interference} more generally as $Y^{s_t \overline{\bm{s}}_t} = g_{s_t \overline{\bm{s}}_t} ( \bm{W}^{ex}_t, \calE_t )$. We also redefine the conditional PTE more generally as the PTE conditional on the {\sl potential treatment history} $\overline{\bm{s}}_t$; i.e., $\delta^\text{PTE}_t \big( \overline{\bm{s}}_t \big) = Y^{1 \overline{\bm{s}}_t}_t - Y^{0 \overline{\bm{s}}_t}_t$.

Let $\overline{\bm{X}}_t$ denote the {\sl observed treatment history} wherein $X$ is first received or assigned at $t = 1$. Following Sävje et al (2021) \cite{2021_savje_etal}, we define the general formula for the {\sl historical PTE} as $\delta_t \big( \overline{\bm{x}}_t \big) = Y^{1 \overline{\bm{x}}_t}_t - Y^{0 \overline{\bm{x}}_t}_t$. Hence, the historical PTE is just the value of the conditional PTE for the particular treatment history $\overline{\bm{x}}_t \in \{ \overline{\bm{s}}_t \}$; i.e., $\delta_t \big( \overline{\bm{x}}_t \big) = \sum_{\{ \overline{\bm{s}}_t \}} \delta^\text{PTE}_t \big( \overline{\bm{s}}_t \big) I \big( \overline{\bm{s}}_t = \overline{\bm{x}}_t \big) = \delta^\text{PTE}_t \big( \overline{\bm{x}}_t \big)$.

Following our convention in Subsection \ref{subsec:temporal-considerations-core-concepts} regarding lagged outcomes for autocorrelation, a given quantity need not depend on all elements of $\overline{\bm{X}}_t$. For example, each PO in Table \ref{tab:carrymod} only depends on $\overline{\bm{s}}_t$ through $s_{t-1}$; i.e., $Y^{s_t \overline{\bm{s}}_t}_t = Y^{s_t s_{t-1}}_t$. Hence, the historical PTE is just $\delta_t \big( \overline{\bm{x}}_t \big) = \delta^\text{PTE}_t \big( x_{t-1} \big) = Y^{1 x_{t-1}}_t - Y^{0 x_{t-1}}_t = \beta_X + I(t>1) I( x_{t-1} = 1 ) \beta_{Xco}$.

We define the corresponding {\sl historical APTE} (HAPTE) as $\delta^\text{HAPTE}_{(m)} \big( \overline{\bm{x}}_m \big) = \frac{1}{m} \sum_{t=1}^m \delta_t \big( \overline{\bm{x}}_t \big)$. The HAPTE is the average effect taken over all periods when the observed exposure and outcome histories are $\overline{\bm{x}}_m$ and $\overline{\bm{y}}_m$, respectively. It answers the question, \lq\lq What was the average effect {\sl over my particular history of treatments, endogenous characteristics, and unchangeable exogenous characteristics (e.g., weather)}?" In Table \ref{tab:carrymod}, the HAPTE is $\delta^\text{HAPTE}_{(m)} \big( \overline{\bm{x}}_m \big) = \beta_X + I(m>1) \frac{1}{m} \sum_{t=2}^m I( x_{t-1} = 1 ) \beta_{Xco}$.

The HAPTE is a retrospective quantity that is useful if the main goal is to understand past effects given past behavior---to diagnose what already happened. However, suppose the intent is to change health outcomes by changing behavior (i.e., not replicate the same treatment history). Then understanding the possible {\sl future} effect or impact of a repeatable treatment or intervention becomes the main goal, and the HAPTE may not be the appropriate estimand.

\subsubsection{APTE Derivation}\label{subsec:apte-derivation}

As mentioned at the end of Subsection \ref{subsec:apte-core-concepts}, serial interference cannot generally cannot be ruled out, and effect constancy generally will not hold. To answer our n-of-1 question posed at the end of Subsection \ref{subsec:nof1objective}, we must first generally define each CAPO used to calculate $\delta^\text{PTE}_t = Y^{1 \bigcdot}_t - Y^{0 \bigcdot}_t$. (See the Appendix for the full derivations of all formulas in this Subsection.)

Redefine serial causal consistency from Subsection \ref{subsec:interference} more generally as $Y_t = Y^{X_t \overline{\bm{X}}_t}_t = \sum_{\{ s_t, \overline{\bm{s}}_t \}} Y^{s_t \overline{\bm{s}}_t}_t I ( X_t = s_t, \overline{\bm{X}}_t = \overline{\bm{s}}_t )$. We now generally define the CAPO as $Y^{s_t \bigcdot}_t = Y^{s_t}_t$ at $t = 1$, and otherwise
\begin{align}
    Y^{s_t \bigcdot}_t
        &=
        	E_{\overline{\bm{X}}_t} \left(
        	    Y^{s_t \overline{\bm{X}}_t}_t \Big| X_t = s_t
            \right)
        =
        	\sum_{\{ \overline{\bm{s}}_t \}}
        	    Y^{s_t \overline{\bm{s}}_t}_t
        	    \Pr \left( \overline{\bm{X}}_t = \overline{\bm{s}}_t \big| X_t = s_t \right)
        \label{eqn:capo}
\end{align}
using the serial causal consistency formula. Note that $\delta^\text{PTE}_t$ is similar to the \lq\lq average contemporary direct effect" of Wang (2021) \cite{2021_wang}. However, whereas Wang takes the expectation over all individuals at a given period, the expectation in equation \eqref{eqn:capo} is taken over all periods prior to $t$ for a single individual.

Suppose $X$ is randomized at every period, as in an n-of-1 experiment. This reduces the conditional probability component of equation \eqref{eqn:capo} to $\Pr \big( \overline{\bm{X}}_t = \overline{\bm{s}}_t \big) = \prod_{s_k \in \overline{\bm{s}}_t} \Pr \big( X_k = s_k \big)$, and the CAPO to $Y^{s_t \bigcdot}_t = E_{\overline{\bm{X}}_t} \Big( Y^{s_t \overline{\bm{X}}_t}_t \Big)$. Recalling equation \eqref{eqn:apte_simple}, we have the {\sl APTE for an observation period of length $m$},
\begin{align}
    \delta^\text{APTE}_{(m)}
        &=
            E^{(m)} \left(
                \delta^\text{PTE}_t
            \right)
        =
            \frac{1}{m} \sum_{t=1}^m
            E \left\{
                \delta^\text{PTE}_t \left( \overline{\bm{X}}_t \right)
            \right\}
        =
            \frac{1}{m} \sum_{t=1}^m
            E_{\calE_t} \left\{
                \delta_t ( \bigcdot )
            \right\}
        =
            E \left\{
                \delta_t ( \bigcdot )
            \right\}
        =
            \delta^\text{HAPTE}_{(m)}
        \label{eqn:apte}
,
\end{align}
by randomization of $X_t$ and the historical PTE and HAPTE definitions. We see that this APTE is equal to the mean HAPTE taken over this same observation period, $\delta^\text{HAPTE}_{(m)}$, taken over all possible randomized exposure histories. Sävje et al (2021) \cite{2021_savje_etal} call the third term an {\sl average distributional shift effect}---a generalization of the original group- and population-level quantities in Hudgens and Halloran (2008) \cite{2008_hudgens_halloran}.

The n-of-1 experimental quantity of equation \eqref{eqn:apte} answers the question posed at the end of Subsection \ref{subsec:nof1objective}. This APTE can represent or approximate an average effect over a future time interval with exogenous characteristics identically distributed to those observed during the experiment; i.e., with the same joint distribution of $\overline{\bm{W}}^{ex}_{m+1} = \big( \bm{W}^{ex}_1, \hdots, \bm{W}^{ex}_m \big)$. We can estimate this APTE using only observed data (i.e., without knowing any counterfactuals) because the conditional mean observed outcome at a given period when $X$ is randomized at every period is equal to the mean CAPO at $t$: $E( Y_t | X_t = s_t ) = E ( Y^{s_t \bigcdot}_t )$.

If there is no carryover, autocorrelation, or any other source of serial interference (e.g., an unobserved exogenous confounder $W^{ex}_{t-2}$ that affects both $X_{t-1}$ and $Y_t$), then the treatment history does not influence the POs. That is, $Y^{s_t \overline{\bm{s}}_t}_t = Y^{s_t}_t$ and $\delta_t \big( \overline{\bm{x}}_t \big) = Y^1_t - Y^0_t$. This is the case in Table \ref{tab:carrymod} when $\beta_{Xco} = 0$, which also implies that $\delta_t \big( \overline{\bm{x}}_t \big) = \beta_X$ is effect-constant over all $t$. This in turn implies a constant mean HAPTE, and hence, a constant APTE; i.e., $\delta^\text{APTE}_{(m)} = \delta^\text{HAPTE}_{(m)} = \beta_X$ for any $m$.

\section{Estimation Methods}\label{sec:methods}

In an n-of-1 experiment, randomization eliminates confounding due to autocorrelation and carryover. There is no such guarantee in an n-of-1 observational study; the possibility of confounding cannot be ignored. Thankfully, a number of methods exist to address or adjust for confounding in nomothetic studies, thereby to estimate the ATE.

\subsection{Two Common Approaches}\label{subsec:gfstats}

\subsubsection{G-Formula}\label{subsec:g-formula}

Recall the notation and concepts from Section \ref{subsec:ate}. To estimate the ATE, we can estimate $E \big( Y^x \big)$ directly if $X$ is randomized by estimating $E ( Y | X = x )$ using only observed values. We need not directly model the outcome mechanism $Y = g ( X, \bm{W}, \calE )$ using an {\sl outcome model}. This follows from the law of total expectation.

Let $R=1$ denote the case when $X$ is randomized, and $R=0$ otherwise. Rewrite the outcome model as $g ( X, \bm{W}, \calE, R )$. Note that the expectation of the outcome model when $X = x$, $\bm{W} = \bm{w}$, and $R = r$ is $E_\calE \big( g ( x, \bm{w}, \calE, r ) \big) = E ( Y | X = x, \bm{W} = \bm{w}, R = r )$. By the law of total expectation, we have:
\begin{align}
    E \left( Y | X = x, R = 1 \right)
        &= E_{\bm{W}} \left\{ E \left( Y | X = x, \bm{W}, R = 1 \right) | X = x, R = 1 \right\} \nonumber \\
        &= E_{\bm{W}} \left\{ E \left( Y | X = x, \bm{W}, R = 1 \right) | R = 1 \right\} \text{ because } X \text{ is randomized} \nonumber \\
        &= E \left( Y^x \right) \text{ by \eqref{eqn:porand}}
        \label{eqn:eyxformula}
\end{align}
This result shows why we can infer $E \big( Y^x \big)$ from our estimate $\hat{E} ( Y | X = x, R = 1 )$ without having to model $g ( X, \bm{W}, R, \calE )$.

If $X$ is not randomized, is there a way to use \eqref{eqn:eyxformula} to estimate $E \big( Y^x \big)$? The surprising answer is yes, via the key insight of a well-established method called the {\sl g-computation formula} or simply {\sl g-formula} (a.k.a. direct standardization, stratification, regression adjustment, and the back-door adjustment formula) \cite{1986_robins, 1995_pearl_robins, 2004_lunceford_davidian, 2006-hernan-robins, 2009_robins_hernan, 2009_pearl, 2014_morgan_winship}. The g-formula is the implied result of changing the equivalence between the penultimate and final terms in \eqref{eqn:porand} to $E_{\bm{W}} \big\{ E \big( Y | X = x, \bm{W}, R = 0 \big) | R = 0 \big\} = E \big( Y^x \big)$ whenever {\sl randomization invariance} holds \cite{2018_daza}.

Randomization invariance consists of two assumptions. {\sl Data-generation invariance} (DGI) holds when $g ( X, \bm{W}, \calE, R ) = g ( X, \bm{W}, \calE )$. {\sl Distributional invariance} (DI) holds when $\bm{W} \indep R$. Together, DGI and DI imply
\begin{align*}
    E_{\bm{W}} \left\{ E \left( Y | X = x, \bm{W}, R = 0 \right) | R = 0 \right\}
        &= E_{\bm{W}} \left\{ E \left( Y | X = x, \bm{W} \right) | R = 0 \right\} \text{ by DGI} \nonumber \\
        &= E_{\bm{W}} \left\{ E \left( Y | X = x, \bm{W} \right) | R = 1 \right\} \text{ by DI} \nonumber \\
        &= E \left( Y^x \right) \text{ by \eqref{eqn:eyxformula}}
.
\end{align*}
The result is the g-formula, $E_{\bm{W}} \big\{ E \big( Y | X = x, \bm{W} \big) \big\} = E \big( Y^x \big)$.

G-formula estimation often proceeds as follows. First, fit the outcome model $g ( X, \bm{W}, \calE, R = 0 )$ (or mathematically equivalent PO mechanism $g_X ( \bm{W}, \calE, R = 0 )$ from Subsection \ref{subsec:causal-inference-core-concepts}) to the observed data. DGI implies that these outcomes are generated in exactly the same way regardless of whether or not $X$ is randomized. Interestingly, DGI can be used to show that the {\sl do operator} introduced by Pearl \cite{2009_pearl} is a more general case of the joint condition $\{ X=x, R=1 \}$. That is, randomization of $X$ is a particular way to set or fix $X$, such that $p ( y | do(x) ) = p ( y | X = x, R = 1 )$.

Next, use the fitted model to estimate the mean outcomes $E ( Y | X = s, \bm{W}, R = 0 )$. In machine learning parlance, predict the outcomes using $Y \sim f( X = s, \bm{W}, R = 0 )$. Our notational use of $s$ versus $x$ is deliberate, and highlights this crucial step in g-formula estimation. It conveys that we cannot use the observed $i = 1, \dots, n$ sample values $\big( x_1, \dots, x_n \big)$ for prediction because they may be correlated with the observed distribution of all other causes and contextualizers $( \bm{w}_1, \dots, \bm{w}_n )$. How then do we proceed?

We wish to replicate the setting wherein $R = 1$. By DI, we do not need to modify the observed distribution of $( \bm{w}_1, \dots, \bm{w}_n )$. DI implies that the latter are similarly distributed regardless of whether or not $X$ is randomized (e.g., in both an RCT and in the real world of everyday life). In Subsection \ref{subsec:idiographic-gformula}, we will characterize the transportability of an APTE estimated using model-twin randomization in terms of partial DI by partitioning $\bm{W}$.

Now recall that $X \indep \bm{W}$ when $R = 1$. Hence, we must modify the sample distribution of $\big( X_1, \dots, X_n \big)$ such that $X_i \indep \bm{W}_{i'}$ where $i \ne i'$ for all $i$ and $i' = 1, \dots, n$. One way to do this is to first randomly generate a new set of values $\big( X^{R=1}_1, \hdots, X^{R=1}_n \big)$ independent of $( \bm{w}_1, \dots, \bm{w}_n )$ using an appropriate distribution (e.g., Bernoulli for binary $X$), then predict $Y_i$ for $X^{R=1}_i$. Another option is to predict every PO corresponding to each value of $s$ for each study participant; e.g., $\big( \big\{ Y^0_1, Y^1_1 \big\} \hdots, \big\{ Y^0_n, Y^1_n \big\} \big)$ for binary $X$ with support $s = 0, 1$. The final step is to estimate $E \big( Y^s \big)$ by marginalizing the predicted outcomes over $( \bm{w}_1, \dots, \bm{w}_n )$. We thereby estimate the ATE as a difference between the estimated mean POs; e.g., $\hat{E} \big( Y^1 \big) - \hat{E} \big( Y^0 \big)$.

To illustrate, recall our early example in Subsection \ref{subsec:causal-inference-core-concepts} with $g ( X, \bm{W}, \calE, R = 0 ) = \beta_0 + \beta_1 X + \bm{W} \bm{\beta}_2 + \calE$. First estimate the parameters $\{ \beta_0, \beta_1, \bm{\beta}_2 \}$. Then for every study participant $i = 1, \hdots, n$, either predict outcome values as $\hat{Y}_i$ based on $X^{R=1}_i$, or predict both POs $\big\{ \hat{Y}^0_i, \hat{Y}^1_i \big\}$. Finally, estimate $E \big( Y^s \big)$ as $\hat{E} \big( Y^s \big) = \sum_{i=1}^n \hat{Y}_i I \big( X^{R=1}_i = s \big) \big/ \sum_{i=1}^n I \big( X^{R=1}_i = s \big)$ if using randomly generated $X$ values, or as $\hat{E} \big( Y^s \big) = \frac{1}{n} \sum_{i=1}^n \hat{Y}^s_i$ if using predicted POs. Estimate the ATE as $\hat{E} \big( Y^1 \big) - \hat{E} \big( Y^0 \big)$.

The g-formula approach is particularly appealing for two reasons. It requires fitting an outcome model---an extremely common and often default practice in both statistics and machine learning; explicitly in the former. The approach is also experimentally intuitive: If we have observed all confounders and know the outcome mechanism, then we can correctly estimate the ATE by replicating the probability conditions of a corresponding hypothetical RCT.

\subsubsection{Propensity Score}\label{subsec:propensity-score}

What if we do not know enough about how to model the outcome? Suppose we can instead specify a reasonably correct {\sl exposure or propensity model} for $\Pr ( X = s | \bm{W} )$; i.e., the probability that the exposure $X = s$ for a given level of $\bm{W}$. Then we can use a number of {\sl propensity score} methods \cite{1983_rosenbaum_rubin, 2001_hirano_imbens, 2004_lunceford_davidian} to estimate the ATE. These include the matching and inverse probability weighting (IPW) approaches mentioned in the Appendix. In this approach, $X = I \big( \calE^X > \Pr ( X = 1 | \bm{W} ) \big)$ where $\calE^X$ is uniformly distributed between 0 and 1. This technique is inspired by Horvitz-Thompson weights used in survey sampling \cite{1952-horvitz-thompson, 1997-robins}.

Our main focus will be to characterize the performance (with respect to empirical bias) of a g-formula estimation approach. However, we will also propose a complementary propensity-score IPW method for comparison. We will not explore {\sl doubly robust} methods (e.g., \lq\lq augmented IPW"), which involve specifying both outcome and propensity models. They are called as such because only one of these two models needs to be specified correctly to guarantee consistent estimation \cite{2005-bang-robins}.

\subsubsection{Modeling Flexibility}\label{subsec:modflex}

Now suppose the outcome and propensity mechanisms are either unknown or cannot otherwise be reasonably justified (e.g., due to a lack of accumulated theoretical evidence). This is the case in our setting of causal hypothesis generation (as well as the field of causal discovery): Our primary objective is to {\sl identify}, {\sl select}, or {\sl propose} plausible models for a small set of posited causal effects, rather than {\sl estimate} effects and associations posited by scientifically defensible or otherwise well-established a priori models \cite{2010_arlot_celisse, 2007_yang}.

Note that the outcome mechanism in equation \eqref{eqn:eyxformula} can take any appropriate functional form that can meet all relevant assumptions for causal identification. These have traditionally been linearized parametric or semi-parametric models in the statistics literature. However, equation \eqref{eqn:eyxformula} applies to the true mechanism, which may or may not be a linearized model. Hence, we can also use supervised learning methods that allow us to fit models focused on characterizing the relationship between $X$, the exposure of interest, and the outcome $Y$, while also flexibly accommodating non-exposure variables $\bm{W}$. Likewise holds for modeling the propensity of $X$.

The decision tree method is one such non-parametric approach, that Athey and Imbens (2015) \cite{2015_athey_imbens} used to estimate a conditional ATE for a continuous outcome (specified as a difference between mean POs). Their tree-based approach can be used to estimate an APTE if conditional WSS holds (i.e., the outcomes are WSS conditional on the causes). WSS replaces the assumption of conditional independence (i.e., the outcomes are mutually independent conditional on the causes) needed for consistent effect estimation. Unlike linearized models that require prespecification of interaction terms, the tree-based method known as random forests (RF) implicitly allows for multiple interactions between predictors in predicting the outcome.

Following Athey and Imbens (2015) \cite{2015_athey_imbens}, we will take a Single Tree approach, and model the outcome as a function of both the exposure and other causes using RF. For comparison, their Two Trees approach for conducting feature selection (i.e., for non-exposure causes important for predicting potential outcomes) involves fitting separate outcome models for each exposure level. To model the exposure propensity, we will simply model the propensity directly (versus their more sophisticated Transformed Outcome Tree approach).

\subsection{Idiographic Approaches}\label{subsec:idiographic-approaches}

In this Subsection, we will define an idiographic g-formula \cite{2018_daza} for estimating the APTE defined in equation \eqref{eqn:apte}. We will address the simple case in which all associations between the outcome and its predictors are both stable and period-stable.

We will also provide a sketch of an IPW approach to complement MoTR, called propensity score twin (PSTn, pronounced \lq\lq piston"). Unlike MoTR, PSTn is not a sequential Monte Carlo method. Instead, PSTn only predicts the exposure probability at each period once after modeling the probability of exposure; i.e., $\Pr \big( X_t = s | \bm{W}_t \big)$. The general procedure involves fitting the propensity model $\pi_t = \Pr \big( X_t = 1 | \bm{W}_t \big)$ to the data. Each observed $Y_t$ is then weighted by the reciprocal of its corresponding estimated propensity based on the observed exposure $x_t$. For $s \in \{ 0, 1 \}$, $E \big( Y^s \big)$ is estimated as the average of the weighted outcomes. See the Appendix for full details on the PSTn procedure.

Following Subsection \ref{subsec:modflex}, we will also implement more flexible RF-based analogues to each of these two approaches. Because estimation is the focus of our paper, we will only characterize the bias of the PSTn method.

\subsubsection{Idiographic G-Formula and APTE Transportability}\label{subsec:idiographic-gformula}

A critical feature of equation \eqref{eqn:apte} is that it in fact describes a {\sl partially transportable APTE}, defined below. We will start by making the tacit assumptions in the relevant SCMs and PO mechanisms explicit, and thereby understand what they imply about transportability of the APTE. (See the Appendix for the full derivations of all formulas in this Subsection.)

First note that in the most general case, at least one element of $\overline{\bm{X}}_t$ is a cause of at least one element of $\bm{W}^{en}_t$. In Subsection \ref{subsec:deriving-the-apte}, we considered the SCM example $Y_t = \beta_0 + \beta_X X_t + \beta_{ar} Y_{t-1} + \bm{W}^{ex}_t \bm{\beta}_{ex} + \calE_t$ for $t>1$. We can concisely write the corresponding PO mechanism using the notation introduced in Subsection \ref{subsec:historical-effects} as $Y^{s_t \overline{\bm{s}}_t}_t = \beta_0 + \beta_X s_t + \beta_{ar} Y^{\overline{\bm{s}}_t}_{t-1} + \bm{W}^{ex}_t \bm{\beta}_{ex} + \calE_t = g_{s_t \overline{\bm{s}}_t} \big( \bm{W}^{ex}_t, \calE_t, \overline{\bm{R}}_t \big)$ where $\overline{\bm{R}}_t$ is a vector of randomization indicators. This makes it clear that $Y^{s_t \overline{\bm{s}}_t}$ in equation \eqref{eqn:capo} depends on $\bm{W}^{ex}_t$ and $\overline{\bm{R}}_t$ such that the explicit CAPO expression is
\begin{align}
    Y^{s_t \bigcdot}_t
        &=
        	E_{\overline{\bm{X}}_t} \left(
        	    Y^{s_t \overline{\bm{X}}_t}_t \big| X_t = s_t, \bm{W}^{ex}_t, \overline{\bm{R}}_t
            \right)
        =
        	\sum_{\{ \overline{\bm{s}}_t \}}
        	    Y^{s_t \overline{\bm{s}}_t}_t
        	    \Pr \left( \overline{\bm{X}}_t = \overline{\bm{s}}_t \big| X_t = s_t, \bm{W}^{ex}_t, \overline{\bm{R}}_t \right)
        \label{eqn:capo_explicit}
.
\end{align}

Now suppose $X$ is randomized at every period; i.e., $\overline{\bm{R}}_{m+1} = \bm{1}$. This reduces the conditional probability component of equation \eqref{eqn:capo_explicit} to $\Pr \big( \overline{\bm{X}}_t = \overline{\bm{s}}_t | \overline{\bm{R}}_t = \bm{1} \big) = \prod_{s_k \in \overline{\bm{s}}_t} \Pr \big( X_k = s_k | \overline{\bm{R}}_t = \bm{1} \big)$. However, $Y^{s_t \overline{\bm{s}}_t}$ still depends on $\bm{W}^{ex}_t$. Hence, the CAPO is now written as $Y^{s_t \bigcdot}_t = E_{\overline{\bm{X}}_t} \big( Y^{s_t \overline{\bm{X}}_t}_t | \bm{W}^{ex}_t, \overline{\bm{R}}_t = \bm{1} \big)$.

With our new notation, we can explicitly write out the APTE of equation \eqref{eqn:apte}. First, we use equation \eqref{eqn:capo_explicit} to explicitly write $\delta^\text{PTE}_t \big( \overline{\bm{X}}_t, \bm{W}^{ex}_t, \overline{\bm{R}}_t = \bm{1} \big) = E_{\overline{\bm{X}}_t} \big( Y^{1 \overline{\bm{X}}_t}_t | \bm{W}^{ex}_t, \overline{\bm{R}}_t = \bm{1} \big) - E_{\overline{\bm{X}}_t} \big( Y^{0 \overline{\bm{X}}_t}_t | \bm{W}^{ex}_t, \overline{\bm{R}}_t = \bm{1} \big)$, where $\calE$ is implicitly conditioned on in each term. We have:
\begin{align}
    \delta^\text{APTE}_{(m)}
        &=
            E^{(m)} \left(
                \delta^\text{PTE}_t
                \big| \overline{\bm{R}}_t = \bm{1}
            \right)
            \nonumber \\
        &=
            \frac{1}{m} \sum_{t=1}^m
            E_{\overline{\bm{X}}_t, \calE_t} \left\{
                \delta^\text{PTE}_t \left( \overline{\bm{X}}_t, \bm{W}^{ex}_t, \calE_t, \overline{\bm{R}}_t = \bm{1} \right)
                \big| \bm{W}^{ex}_t, \overline{\bm{R}}_t = \bm{1}
            \right\}
            \nonumber \\
        &=
            \frac{1}{m} \sum_{t=1}^m
            E_{\calE_t} \left\{
                \delta_t \left( \bigcdot, \bm{W}^{ex}_t, \calE_t, \overline{\bm{R}}_t = \bm{1} \right)
            \right\}
            \nonumber \\
        &=
            E \left\{
                \delta_t ( \bigcdot )
                \big| \overline{\bm{R}}_{m+1} = \bm{1}
            \right\}
            \nonumber \\
        &=
            \delta^\text{HAPTE}_{(m)}
        \label{eqn:apte_explicit}
\end{align}
As before, we can estimate this APTE using only observed data when $\overline{\bm{R}}_{m+1} = \bm{1}$. Let $E^{(m)} \big( Y^{s_t \bigcdot}_t \big| \overline{\bm{R}}_{m+1} = \bm{1} \big)$ represent the {\sl mean CAPO for an observation period of length $m$}, which is equal to the conditional mean observed outcome at a given period:
\begin{align}
    E \left(
        Y_t
        \big| X_t = s_t, \overline{\bm{R}}_t = \bm{1}
    \right)
        =
            E \left(
                Y^{X_t \overline{\bm{X}}_t}_t \Big| X_t = s_t, \overline{\bm{R}}_t = \bm{1}
            \right)
        =
            E \left(
                Y^{s_t \bigcdot}_t
                \big| \overline{\bm{R}}_{m+1} = \bm{1}
            \right)
        \label{eqn:eytbarxt_nof1}
\end{align}
From \eqref{eqn:apte_explicit}, note that $
E \big\{
    \delta_t ( \bigcdot )
    \big| \overline{\bm{R}}_{m+1} = \bm{1}
\big\}
    =
        E \big(
            Y^{1 \bigcdot}_t -
            Y^{0 \bigcdot}_t
            \big| \overline{\bm{R}}_{m+1} = \bm{1}
        \big)
    =
        E \big(
            Y^{1 \bigcdot}_t
            \big| \overline{\bm{R}}_{m+1} = \bm{1}
        \big) -
        E \big(
            Y^{0 \bigcdot}_t
            \big| \overline{\bm{R}}_{m+1} = \bm{1}
        \big)
$.
 
To begin defining the {\sl idiographic g-formula}, first recall our SCM example, and note that $Y_{t-1}$ is a function of $\overline{\bm{X}}_t$. In general, we can write $\bm{W}^{en}_t$ as a function of $\overline{\bm{X}}_t$; i.e., $\bm{W}^{en}_t \big( \overline{\bm{X}}_t \big)$. This lets us write $Y_t = g \Big( X_t, \bm{W}^{en}_t \big( \overline{\bm{X}}_t \big), \bm{W}^{ex}_t, \calE_t, \overline{\bm{R}}_t \Big)$, or simply $Y_t = g \big( X_t, \overline{\bm{X}}_t, \bm{W}^{ex}_t, \calE_t, \overline{\bm{R}}_t \big)$. We now derive the empirical mean CAPO in terms of the SCM by adapting equation \eqref{eqn:eyxformula} as follows:
\begin{align}
    E \left(
        Y_t
        \big| X_t = s_t, \overline{\bm{R}}_t = \bm{1}
    \right)
        &=
            E^{(m)}_{\bm{W}^{en}_t, \calE_t, \bm{W}^{ex}_t} \left(
                Y_t
                \big| X_t = s_t, \overline{\bm{R}}_t = \bm{1}
            \right)
            \nonumber \\
        &=
            E^{(m)}_{\bm{W}^{ex}_t} \left(
                E_{\calE_t} \left[
                    E_{\overline{\bm{X}}_t} \left\{
                        g \left( s_t, \overline{\bm{X}}_t, \bm{W}^{ex}_t, \calE_t, \overline{\bm{R}}_t = \bm{1} \right)
                        \big| \overline{\bm{R}}_t = \bm{1}
                    \right\}
                \right]
                \big| \overline{\bm{R}}_t = \bm{1}
            \right)
            \nonumber \\
        &=
            E \left(
                Y^{s_t \bigcdot}_t
                \big| \overline{\bm{R}}_{m+1} = \bm{1}
            \right)
            \text{ by \eqref{eqn:eytbarxt_nof1}}
        \label{eqn:eyxformula_nof1}
\end{align}
We thereby define the idiographic g-formula as the implied result of changing the equivalence between the penultimate and final terms in \eqref{eqn:eyxformula_nof1} to $
E^{(m)}_{\bm{W}^{ex}_t} \big(
    E_{\calE_t} \big[
        E_{\overline{\bm{X}}_t} \big\{
            g \big( s_t, \overline{\bm{X}}_t, \bm{W}^{ex}_t, \calE_t, \overline{\bm{R}}_t = \bm{0} \big)
            \big| \overline{\bm{R}}_t = \bm{1}
        \big\}
    \big]
    \big| \overline{\bm{R}}_t = \bm{0}
\big)
=
    E^{(m)} \big(
        Y^{s_t \bigcdot}_t
        \big| \overline{\bm{R}}_{m+1} = \bm{1}
    \big)
$ whenever {\sl sequential randomization invariance} holds.

We define sequential randomization invariance as consisting of two assumptions that directly parallel the non-sequential case. {\sl Sequential data-generation invariance} (SDGI) holds when $g \big( X_t, \bm{W}_t, \calE_t, \overline{\bm{R}}_t \big) = g \big( X_t, \bm{W}_t, \calE_t \big)$. {\sl Sequential distributional invariance} (SDI) holds when $\bm{W}^{ex}_t \indep \overline{\bm{R}}_t$. We have:
\begin{align*}
    & E^{(m)}_{\bm{W}^{ex}_t} \left(
        E_{\calE_t} \left[
            E_{\overline{\bm{X}}_t} \left\{
                g \left( s_t, \overline{\bm{X}}_t, \bm{W}^{ex}_t, \calE_t, \overline{\bm{R}}_t = \bm{0} \right)
                \big| \overline{\bm{R}}_t = \bm{1}
            \right\}
        \right]
        \big| \overline{\bm{R}}_t = \bm{0}
    \right)
    \nonumber \\
        &=
            E^{(m)}_{\bm{W}^{ex}_t} \left(
                E_{\calE_t} \left[
                    E_{\overline{\bm{X}}_t} \left\{
                        g \left( s_t, \overline{\bm{X}}_t, \bm{W}^{ex}_t, \calE_t, \overline{\bm{R}}_t = \bm{1}
                        \right)
                        \big| \overline{\bm{R}}_t = \bm{1}
                    \right\}
                \right]
                \big| \overline{\bm{R}}_t = \bm{0}
            \right)
            \text{ by SDGI}
            \nonumber \\
        &=
            E^{(m)}_{\bm{W}^{ex}_t} \left(
                E_{\calE_t} \left[
                    E_{\overline{\bm{X}}_t} \left\{
                        g \left( s_t, \overline{\bm{X}}_t, \bm{W}^{ex}_t, \calE_t, \overline{\bm{R}}_t = \bm{1}
                        \right)
                        \big| \overline{\bm{R}}_t = \bm{1}
                    \right\}
                \right]
                \big| \overline{\bm{R}}_t = \bm{1}
            \right)
            \text{ by SDI}
            \nonumber \\
        &=
            E \left(
                Y^{s_t \bigcdot}_t
                \big| \overline{\bm{R}}_{m+1} = \bm{1}
            \right)
            \text{ by \eqref{eqn:eyxformula_nof1}}
\end{align*}
The result is the idiographic g-formula:
\begin{align}
    & E^{(m)}_{\bm{W}^{ex}_t} \left(
        E_{\calE_t} \left[
            E_{\overline{\bm{X}}_t} \left\{
                g \left( s_t, \overline{\bm{X}}_t, \bm{W}^{ex}_t, \calE_t
                \right)
                \big| \overline{\bm{R}}_t = \bm{1}
            \right\}
        \right]
    \right)
        =
            E \left(
                Y^{s_t \bigcdot}_t
                \big| \overline{\bm{R}}_{m+1} = \bm{1}
            \right)
        \label{eqn:gformula_idiographic}
\end{align}

Similar to its nomothetic counterpart, estimation of the idiographic g-formula requires a positivity assumption to be met. Define {\sl sequential positivity} as the condition that $p \big( x_t, \overline{\bm{x}}_t, \bm{w}^{ex}_t \big| \overline{\bm{R}}_t = \bm{0} \big) \ne 0$ for all values in the joint support of $\big\{ X_t, \overline{\bm{X}}_t, \bm{W}^{ex}_t \big\}$. This condition could be violated, for example, due to strong effects of past treatments $\overline{\bm{X}}_t$ on treatment $X_t$; either directly, or indirectly through their effects on past outcomes $\overline{\bm{Y}}_t$ that in turn cause $X_t$. One way to check if this assumption is sufficiently met is to examine pairwise cross-correlations between $\big\{ X_t, \overline{\bm{X}}_t, \bm{W}^{ex}_t \big\}$. When fitting a linear model for $g \big( X_t, \bm{W}_t, \calE_t \big)$, one could also check for multicollinearity.

Sequential randomization invariance exactly determines when the APTE is transportable. In particular, the APTE is transportable across settings wherever the SDI assumption holds (i.e., settings with the same distribution of exogenous characteristics). The SDI assumption is a \lq\lq partial" DI assumption in that not all other causes of $Y_t$ need be identically distributed regardless of randomization status. Indeed, the distribution of endogenous causes $\bm{W}^{en}_t$ depends on if or how $\overline{\bm{X}}_t$ is randomized, such that $\bm{W}^{en}_t \indep \overline{\bm{R}}_t$ cannot be true. Hence, $\bm{W}_t \indep \overline{\bm{R}}_t$ cannot hold in general, thus rendering the APTE only {\sl partially transportable}.

The APTE of equation \eqref{eqn:apte_explicit} is a type of {\sl expected average treatment effect} as defined in Sävje et al (2021) \cite{2021_savje_etal}. To paraphrase the authors, the APTE \lq\lq captures the expected effect of changing a random period's treatment if the current study had been an n-of-1 experiment, with its particular history of unchangeable, exogenous characteristics." But can we define an APTE that is somehow stable over time, and not just specific to a given observation period?

Recall from the end of Subsection \ref{subsec:apte-core-concepts} that effect constancy (i.e., a constant PTE over all periods) may be overly optimistic to expect from real-world observational data. But if the variation in PTEs is bounded, the implied APTE may be stable or constant {\sl in the long run} (i.e., for very long multivariate time series). Equation \eqref{eqn:apte_explicit} allows us to define the {\sl long-run APTE} as
$
    \lim_{m \to \infty} \delta^\text{APTE}_{(m)}
        =
            \delta^\text{APTE}
$, a quantity that exhibits a property we will call {\sl long-run effect constancy}. In the next Section, we will see that it is possible to use a large-enough sample (i.e., long-enough multivariate time series) to accurately estimate an identifiable long-run APTE that is long-run effect-constant.

\subsubsection{Autoregressive Carryover Model and Finite Time Horizon}\label{subsec:arco}

We now define an outcome model (see Subsection \ref{subsec:g-formula}) that we will call the {\sl autoregressive carryover} (ARCO) model, an identifiable linear model that accounts for both carryover and autocorrelation. This model is based on the dynamic regression models of Vieira et al (2017), Kravitz et al (2014), and Schmid (2001) \cite{2017-vieira-etal, 2014_kravitz_etal, 2001_schmid}. The ARCO model lets us derive some useful long-run averages---including a long-run APTE---when WSS holds.

The ARCO model can be thought of as a Granger model because it conceptually resembles models used to assess \lq\lq Granger causality" \cite{1969_granger, 1980_granger, 1988_granger, 2001_kaminski_etal, 2017_lu_etal, 2010_white_lu}. The latter generally describes statistical associations between two or more time series, not causality as defined in the causal inference literature. We define the general ARCO model as
\begin{equation}
    Y_t
        =   \beta_0 +
            \beta_X X_t +
            \overline{\bm{X}}_t^{\ell^X} \bm{\beta}_{co} +
            \bm{X}^\otimes_t \bm{\beta}_{Xco} +
            \overline{\bm{Y}}_t^{\ell^Y} \bm{\beta}_{ar} +
            \bm{Y}^\otimes_t \bm{\beta}_{Xar} +
            \bm{W}^{ex}_t \bm{\beta}_{ex} +
            \calE_t
	\label{eqn:arco}
.
\end{equation}
Here, $\bm{X}^\otimes_t$ is the vector of all unique two-way interactions (i.e., products) between the elements of $\big( X_t, \overline{\bm{X}}_t^{\ell^X} \big)$, and $\bm{Y}^\otimes_t$ is the vector of all unique two-way interactions between the elements of $\big( X_t, \overline{\bm{Y}}_t^{\ell^Y} \big)$.

At least two assumptions must hold in order to estimate a long-run APTE generated by \eqref{eqn:arco}. The first assumption is that $Y_t$ has a small, finite number of endogenous causes; i.e., $\ell^X <<< \infty$ and $\ell^Y <<< \infty$. We will call this the {\sl finite endogeneity} assumption. The second is that $\big\{ ( Y_t ) \big\}$ must be stationary. From Subsection \ref{subsec:temporal-considerations-core-concepts}, $\big\{ ( Y_t ) \big\}$ is stationary when $| \bm{\beta}_{ar} | < 1$, $| \bm{\beta}_{Xar} | < 1$, and $\big\{ ( \bm{W}^{ex}_t ) \big\}$ is stationary. This stationarity assumption can be tested, for example, via the Augmented Dickey Fuller (ADF) and Kwiatkowski-Phillips-Schmidt-Shin (KPSS) unit-root tests, as done in Daza (2018) \cite{2018_daza}. Together, these two assumptions enable the estimation of causal effects using a finite observation period. Hence, we will collectively call them the {\sl finite time horizon} assumption.

Equation \eqref{eqn:arco} can be considered both a special case and generalization of a vector autoregressive model. It is a special case because it involves only three series (one of which is exogenous), but also a generalization because of the interaction terms. The ARCO model can also be generalized as part of the exponential family by treating its linear component as $\eta$, the linear predictor of a generalized linear model (GLM). Specifically, let
$\eta_{ARCO} =
    \beta_0 +
    \beta_X X_t +
    \overline{\bm{X}}_t^{\ell^X} \bm{\beta}_{co} +
    \bm{X}^\otimes_t \bm{\beta}_{Xco} +
    \overline{\bm{Y}}_t^{\ell^Y} \bm{\beta}_{ar} +
    \bm{Y}^\otimes_t \bm{\beta}_{Xar} +
    \bm{W}^{ex}_t \bm{\beta}_{ex}
$. In this conceptualization, dynamic regression models can be considered a special case of the resulting {\sl generalized ARCO} model, with a binary outcome and the canonical logit link function.

Throughout the rest of this paper, we will specify our default SCM $g \big( X_t, \overline{\bm{Y}}_t^{\ell^Y}, \overline{\bm{X}}_t^{\ell^X}, \bm{W}^{ex}_t, \calE_t \big)$ using the ARCO model. We will assume $\big\{ ( Y_t ) \big\}$ and $\big\{ ( \bm{W}^{ex}_t ) \big\}$ are at least WSS unless otherwise stated.

\subsubsection{Order-1 Model with Randomized Treatments}\label{subsec:order-1-model}

Hereafter, we will only consider the order-1 ARCO model $Y_t = \beta_0 + \beta_X X_t + \beta_{co} X_{t-1} + \beta_{Xco} X_t X_{t-1} + \beta_{ar} Y_{t-1} + \beta_{Xar} X_t Y_{t-1} + \bm{W}^{ex}_t \bm{\beta}_{ex} + \calE_t$. Suppose $X$ is randomized with constant probability $\pi = \Pr( X = 1 )$ in an n-of-1 experiment. Applying the long-run APTE of equation \eqref{eqn:apte_explicit} to this order-1 model yields $\delta^\text{APTE} = \beta_X + \beta_{Xco} \pi + \beta_{Xar} \mu_Y$. For completeness, in an n-of-1 experiment we have the long-run mean outcome
\begin{equation*}
\mu_Y = \frac{\beta_0 + \beta_X \pi + \beta_{co} \pi + \beta_{Xco} \pi^2 + \bm{\mu}_{ex} \bm{\beta}_{ex}}{1 - \beta_{ar} - \beta_{Xar} \pi}
,
\end{equation*}
where $\bm{\mu}_{ex} = E \big( \bm{W}^{ex}_t \big)$. See the Appendix for the derivation.

Suppose the APTE is not modified by carryover such that $\beta_{Xco} = 0$, and is not modified by past outcomes such that $\beta_{Xar} = 0$. Then $\delta^\text{APTE}$ is just the impact of $X_t$ through $\beta_X$. With a positive carryover influence of $\beta_{Xco} > 0$, the impact of $X_t$ is amplified in proportion to the randomization probability $\pi$; a negative influence dampens its impact.

\section{Model-Twin Randomization}\label{subsec:motr}

Model-twin randomization mimics an n-of-1 experiment by randomizing the originally observed sequential exposure $\overline{\bm{x}}_{m+1}$, thereby changing it from a possibly endogenous variable (i.e., that might be affected by past values of $Y$) into an exogenous one unaffected by the outcome. Hence, MoTR is a Monte Carlo method that provides numerically approximate calculations of statistically consistent estimates of the APTE, along with approximate confidence intervals (CIs) for conducting inference. It mirrors the g-formula estimation approach used for fitting marginal structural models \cite{2024-loh-etal} (see Subsection \ref{subsec:statistical-causal-inference}).

\subsection{MoTR Procedure}\label{subsec:motr-procedure}

The general MoTR procedure or algorithm is as follows, whereby the outcome $Y_t$ is sequentially generated.

\begin{enumerate}

    \item Fit the outcome model $\mu_t = E \big( Y_t | X_t, \bm{W}_t \big)$ to the observed data $\big\{ \overline{\bm{y}}_{m+1}, \overline{\bm{x}}_{m+1}, \overline{\bm{w}}_{m+1} \big\}$. The estimator $\hat{\mu}_t$ from fitting the model is the participant's {\sl model twin}---a digital twin \cite{2020-schwartz-etal} that represents the participant's outcome mechanism for estimating the APTE.
    
        \begin{enumerate}
        
            \item If needed, first transform the original values $Y^*$ to $Y$. For example, $Y = \log ( Y^* )$ can change $Y^*$ into a less skewed, more normally distributed variable amenable to adding Gaussian noise to model predictions in a subsequent step below.
            
        \end{enumerate}
        
    \item Run the model twin through a simulated n-of-1 experiment by \lq\lq running the MoTR" as follows. For run (i.e., iteration) $r$:
    
        \begin{enumerate}
        
            \item Randomly permute or shuffle all observed $x_t$. Let $\overline{\bm{X}}_{r,m+1} = \big\{ \big( X_{rt} \big) \big\} = \big( X_{r1}, \hdots, X_{rm} \big)$ represent this randomized sequence. This step preserves the original ratio of exposures to non-exposures, reflecting the exposure's observed overall propensity.
            
            \item Generate $\bm{W}_{rt}$ at each period $t$ sequentially as follows. We need to index this matrix by its run number $r$ because $\bm{W}^{en}_{rt} \in \bm{W}_{rt}$ will change with each run, as detailed below.
            
                \begin{enumerate}
                
                    \item For all exogenous variables, set $\bm{W}^{ex}_{rt} \equiv \bm{w}^{ex}_t$. Take care not change these values at any point while implementing the MoTR procedure, as they represent the observed particular history of unchangeable, exogenous characteristics.
                    
                    \item Set all lags of $Y_t$ contained in $\bm{W}_t$ equal to their values generated in past runs. For example, set $Y_{t-1}$ equal to $\hat{Y}_{r(t-1)}$ as defined in step (c)iii. below.
                    
                    \item Set all lags of $X_t$ contained in $\bm{W}_t$ equal to their corresponding values in $\overline{\bm{X}}_{r,m+1}$.
                    
                \end{enumerate}
                
            \item Generate $Y_{rt}$ at each period $t$ sequentially as follows.
            
                \begin{enumerate}
                
                    \item Predict the outcome as $\hat{\mu}_{rt}  = \hat{E} \big( Y_t | X_t = X_{rt}, \bm{W}_t = \bm{W}_{rt} \big)$.
                    
                    \item Generate random noise $\varepsilon_{rt} \sim N(0, \sigma_\varepsilon)$, where $\sigma_\varepsilon$ is the standard deviation of the residuals from fitting the model to the original data (i.e., $\overline{\bm{y}}_{m+1} - \overline{\hat{\bm{\mu}}}_{m+1}$) (treated as fixed), and $\hat{\mu}_t  = \hat{E} \big( Y_t | X_t = x_t, \bm{W}_t = \bm{w}_t \big)$ depends on the observed data $\big\{ \overline{\bm{x}}_{m+1}, \overline{\bm{w}}_{m+1} \big\}$ for all $\hat{\mu}_t \in \overline{\hat{\bm{\mu}}}_{m+1}$ (i.e., rather than depending on $\big\{ \overline{\bm{X}}_{r,m+1}, \overline{\bm{W}}_{r,m+1} \big\}$).
                    
                    \item Add the random noise to each predicted value as $\hat{Y}_{rt} = \hat{\mu}_{rt} + \varepsilon_{rt}$. This step preserves summary information about residual variation in the outcome, as is commonly done to create prediction intervals in statistics. Note that $\hat{Y}_{rt}$ may be an element of $\bm{W}_{rt'}$ for some future period $t' > t$ as in step (b)iii. above.
                    
                \end{enumerate}
                
            \item For $s \in \{ 0, 1 \}$, estimate $E ( Y^s )$, the mean PO, as the average of the noisy predicted outcomes; i.e., $\hat{E} ( Y^s )_r = \frac{1}{m_s} \sum_{t=1}^m \hat{Y}_{rt} I \big( X_{rt} = s \big)$ where $m_s = \sum_{t=1}^m I \big( X_{rt} = s \big)$. Note that $m_s$ is constant regardless of $r$. Estimate the APTE as $\hat{\delta}^\text{MoTR}_r = \hat{E} ( Y^1 )_r - \hat{E}_r ( Y^0 )_r$.
            
            \item Estimate the corresponding CI likewise; e.g., do this using the common t-test if the predicted outcomes are fairly normally distributed.
            
            \item Calculate the cumulative average APTE over all previous runs as the average (mean or median) of all estimates $\hat{\delta}^\text{MoTR}_r$ up to and including the current run $r$; i.e., $\hat{\delta}^\text{MoTR}_r = \frac{1}{r} \sum_{h=1}^r \hat{\delta}^\text{MoTR}_h$. Calculate the cumulative average CI likewise.
            
        \end{enumerate}
        
    \item Repeat steps 2(a)-2(e) until all three cumulative averages converge based on reasonable criteria. To ensure some stability in cumulative estimates and protect against outlier estimates, consider setting a minimum $r_{min}$ such that $1 <<< r_{min} \le r$). To save on computation time, consider setting a maximum $r_{max}$ such that $r \le r_{max}$.
    
    \item Report the final values from step 3 as the APTE estimate $\hat{\delta}^\text{MoTR}$ along with its CI.

\end{enumerate}

\noindent
See the Appendix for two algebraic identities that can be used to speed up MoTR calculations.

We acknowledge that a CI calculated from a t-test will generally be smaller than the true CI, and therefore provide worse than nominal coverage. However, the statistical goal of this paper is to characterize estimation rather than inference. The reader should explore approaches to constructing CIs that better achieve nominal coverage like the randomization-based CIs derived by Malenica et al (2021, 2023) \cite{2021-malenica-etal, 2023-malenica-etal} and Liang and Recht (2023) \cite{2023-liang-recht}. In particular, see Section {\sl 4. Confidence Sequences} of Malencia et al (2023) \cite{2023-malenica-etal}.

\subsection{Related Approaches}\label{subsec:related-approaches}

\subsubsection{Statistical Causal Inference}\label{subsec:statistical-causal-inference}

Balandat (2016) \cite{2016_balandat} took a similar approach towards causal prediction of electricity consumption using a prognostic-score technique \cite{2008_hansen}, and developed the large-sample theory needed to conduct inference on the APTE if exposure effects do not carry over from one period to the next. Neto et al (2016) \cite{2016_neto_etal} and Schmiedek and Neubauer (2019) \cite{2020-schmiedek-neubauer} both introduced innovative n-of-1 approaches based on instrumental variables and randomized treatments.

In particular, if randomization invariance holds, then randomization status is an instrumental variable that satisfies the exclusion restriction. That is, randomization status is associated with the exposure, but not with any confounders, the outcome, or variation in the outcome not explained by the exposure or confounders. Neto et al (2016, 2017) \cite{2016_neto_etal, 2017_neto_etal} exploited this property of randomization to try to estimate single-person crossover personalized average causal effects.

Cheung et al (2017) \cite{2017_cheung_etal} set the stage for comparing nomothetic and idiographic modeling of the APTE using tree-based approaches. Burg et al (2017) \cite{2017_burg_etal} accomplished a similar goal through mixed-effects modeling using generalized linear models. Balandat also drew a formal connection between individual- and group-level average treatment effects in his Subsection 4.5. And Rohrer and Murayama (2023) \cite{2023-rohrer-murayama} provided a perspective on within- versus between-persons causal inference using longitudinal data. These authors expand on the basic theoretical sketch made by Robins and Hernán (2016) \cite{2016-robins-hernan} around n-of-1 \lq\lq crossover experiments" that is fully elaborated as the APTE framework of Daza (2018) \cite{2018_daza} and the U-CATE framework of Piccininni et al (2024) \cite{2024-piccininni-etal}.

MoTR estimation of the idiographic g-formula can be thought of as serial g-computation. It is operationally similar to estimation of the time-varying g-formula of marginal structural models (MSMs) \cite{1997-robins, 2000_robins_etal, 2018_daza}. This can be seen by comparing and contrasting the MoTR procedure applied in our Simulation Study in Section \ref{sec:sims} to the MSM g-formula estimation procedure outlined in Loh et al (2024) \cite{2024-loh-etal} limited to the case of one lag.

Other approaches mentioned earlier that are very closely related to MoTR include those of Wang (2021) \cite{2021_wang}, Liang and Recht (2023) \cite{2023-liang-recht}, and Malenica et al (2021, 2023) \cite{2021-malenica-etal, 2023-malenica-etal}. Their differences and similarities to MoTR are explained throughout the text where relevant.

\subsubsection{Health Economics: Microsimulation}\label{subsec:health-economics-microsimulation}

Microsimulation models (MicroMods) were originally developed in health economics as a way to draw population-level inferences by simulating and aggregating individual-level outcomes \cite{2018-schofield-etal}. They have been used in oncology and mental health, and to model cardiovascular disease risk \cite{2018-caglayan-etal, 2024-thomson-etal, 2023-wu-etal}.

Rutter et al (2011) \cite{2011-rutter-etal} provide a relevant overview of MicroMods, quoted and summarized throughout the rest of this Subsection. The general goal of MicroMods to predict trends or compare treatment effectiveness \lq\lq is met by developing models that combine results from randomized controlled trials, epidemiologic studies (e.g., case-control and cohort studies), meta-analyses, and expert [opinions]".

The steps involved in implementing a MicroMod are similar to those of MoTR. First, identify distinct states of the outcome, specify stochastic rules for transitioning between states, and specify the model parameters. MoTR uses a continuous-time model for specifying stochastic rules. Next, simulate the population of interest. When an ARCO model is used in MoTR, this \lq\lq population" is a multivariate time series. Finally, both a MicroMod and the MoTR outcome model must be assessed for validity through sensitivity analyses that change modeling assumptions.

Both the MoTR model and a MicroMod must strike a balance in model complexity. A natural-history-based MicroMod can more fully describe biological mechanisms of action that explain how the outcome will change for a given individual. However, such a model can be hard to fit because of its complexity. An epidemiology-based MicroMod that describes the putative effect of a population-level intervention based on \lq\lq the observable portion of the disease process" is easier to fit. But this simpler model must capture enough observable detail for inferring the ATE needed to make population-level decisions. The MoTR model must strike a similar balance. However, it relies on a sufficient \lq\lq epidemiological" model based only on one individual's own data history to infer the APTE for making person-specific decisions.

Ma et al (2024) \cite{2024-ma-etal} draw a more explicit connection to statistical causal inference. They propose improving a state-transition MicroMod for vaccination health policy by applying a recurrent neural network combined with Q-learning. The latter has recently become popular for characterizing the ability of sequential decision rules to individually improve each participant's health in the MRTs and JITAIs mentioned in Subsection \ref{subsec:within-individual-causal-inference} \cite{2019-menictas-etal}.

\section{Simulation Study}\label{sec:sims}

We now describe how we used simulated data to characterize the empirical bias of the MoTR and PSTn estimators ($\hat{\delta}^\text{MoTR}$ and $\hat{\delta}^\text{PSTn}$, respectively). We characterized this bias when each model was correctly specified as the true data-generating mechanism, which we specified as a GLM (i.e., ARCO for the outcome, logistic for the exposure). We also characterized this bias using naive APTE estimation methods; i.e., when no modeling was done, and when the true models were fit but were not adjusted for confounding by either method.

Finally, we characterized the empirical bias of the RF approaches of Subsection \ref{subsec:modflex}. We expected that RF, while more flexible than linearized models in useful ways, would be be more biased because the relevant prediction function is not identical to the true GLM mechanism.

\subsection{Data-generating Procedure}\label{subsec:sims_dgp}

EJD collected 1374 days of sleep duration and step-count data using a Fitbit Charge\texttrademark{} (models 3, 4, and 6) wrist-worn sensor. This was about 3 years and 9 months of data. Hence, each simulated dataset we generated had four years or $m=1461$ days (time points) of data.

We specified our outcome and exposure of interest as follows. A period was specified as a day. Our outcome was daily total sleep time (TST), defined as the total number of hours asleep per night. Our physical-activity exposure was the daily walking cadence. We chose this rather than daily raw step count because we wanted to understand if increasing physical activity overall---regardless of how much time the participant was awake---led to longer or shorter sleep. However, simply getting more sleep leaves one less time to take steps the following day. Hence, raw step count is often associated with sleep duration \cite{2022-chevance-etal}, regardless of how it might affect sleep duration (i.e., our APTE of interest).

In the standard potential outcomes framework, it is reasonable to dichotomize continuous exposures in order to define clear causal effects between distinct groups or levels. Hence, we dichotomized our exposure into high and low physical activity based on the median walking cadence over all days of observation.  We described an exposure as either \lq\lq fast walking cadence" (or \lq\lq fast cadence") or \lq\lq walking quickly" (i.e., $>$ median), or \lq\lq slow walking cadence" (or \lq\lq slow cadence") or \lq\lq walking slowly" (i.e., $\le$ median). We thereby defined the APTE of interest as the mean potential outcome under a fast cadence minus that under a slow cadence.

We generated each simulated dataset (indexed $h$) using the following procedure, which induces autocorrelation and confounding via endogeneity.

\begin{enumerate}

    \item If $t=1$, generate $Y_{ht}$ as a continuous variable drawn from $\beta_0 + \calE_{ht}$ where $\calE_{ht} \sim N \big( 0, \sigma_\varepsilon \big)$. Generate $X_{ht} \in \{ 0, 1 \}$ as a binomial variable with probability $\pi_1 = \Pr \big( X_1 = 1 \big)$.
    
    \item If $t>1$, for both $s=0$ and $s=1$, generate $Y^s_{ht}$ using the ARCO PO mechanism $Y^s_{ht} = \beta_0 + \beta_X s + \beta_{co} X_{h(t-1)} + \beta_{Xco} s X_{h(t-1)} + \beta_{ar} Y_{h(t-1)} + \beta_{Xar} s Y_{h(t-1)} + \bm{W}^{ex}_{ht} \bm{\beta}_{ex} + \calE_{ht}$. Generate $X_{ht}$ using the propensity mechanism $\logit ( \pi_{ht} ) = \alpha_0 + \alpha_{en} Y_{h(t-1)}$ where $\pi_{ht} = \Pr \big( X_{ht} = 1 | Y_{h(t-1)} \big)$. Here, $\alpha_{en} \ne 0$ induces endogeneity by making $X_{ht}$ depend on $Y_{h(t-1)}$. Note that for $t>1$, $\pi_{ht}$ is a logit-normal random variable, which has no analytic mean (see Appendix).
    
    \item Generate the observed outcome $Y_{ht}$ using causal consistency. That is, generate $Y_{ht} = Y^1_{ht} X_{ht} + Y^0_{ht} \big( 1 - X_{ht} \big)$.

\end{enumerate}
The outcome-model parameters we used were $\beta_0 = 12$, $\beta_X = 1.2$, $\beta_{co} = 0$, $\beta_{Xco} = 0$, $\beta_{ar} = -0.9$, $\beta_{Xar} = 0.1$, $\bm{\beta}_{ex} = \bm{0}$, and $\sigma_\varepsilon = 0.5$. Hence, the PO mechanism was $Y^s_{ht} = 12 + 1.2 s - 0.9 Y_{h(t-1)} + 0.1 s Y_{h(t-1)} + \calE_{ht}$ with $E \big( Y^s_{ht} \big) = 12 - 0.9 \mu_Y + s \big( 1.2 + 0.1 \mu_Y \big)$ and $
\text{Var} \big( Y^s_{ht} \big) = ( 0.81 - 0.19 s ) \sigma^2_Y + 0.25$. The corresponding signal-to-noise ratio was $4 \times \big\{ \text{Var} \big( Y^s_{ht} \big) - E \big( Y^s_{ht} \big)^2 \big\}$.

The propensity-model parameters we used were $\alpha_0 = -0.001$ and $\alpha_{en} = 0.05$. We approximated $\pi_{ht} = 0.59$ using 14,710 rows of simulated data (see Appendix on logit-normal random variable). Note that $Y^s_{ht}$ is stationary because $| \beta_{ar} | \le 1$. From the $\delta^\text{APTE}$ formula in Subsection \ref{subsec:order-1-model}, the true APTE was $1.89$.

\subsection{Analysis Methods and Results}\label{subsec:sims_results}

\setlength{\extrarowheight}{4pt}
\begin{table}
\caption{ \label{tab:sims_methods} Simulation study analysis methods.}
\begin{center}
\scalebox{1}{
\begin{tabular}{ l | l | l | l }
\hline
Method & Estimator & Description & Bias Expected \\
\hline
Raw Comparison &
    $\hat{\delta}^\text{raw}$ &
    $\frac{1}{m_1} \sum_{t=1}^m Y_t I \big( X_t = 1 \big) - \frac{1}{m_0} \sum_{t=1}^m Y_t I \big( X_t = 0 \big)$ &
    yes \\
MoTR-GLM &
    $\hat{\delta}^\text{MoTR-GLM}$ &
    $\hat{\delta}^\text{MoTR}$ after fitting GLM outcome model &
    no \\
MoTR-RF &
    $\hat{\delta}^\text{MoTR-RF}$ &
    $\hat{\delta}^\text{MoTR}$ after fitting RF prediction function &
    some \\
PSTn-GLM &
    $\hat{\delta}^\text{PSTn-GLM}$ &
    $\hat{\delta}^\text{PSTn}$ after fitting GLM propensity model &
    no \\
PSTn-RF &
    $\hat{\delta}^\text{PSTn-RF}$ &
    $\hat{\delta}^\text{PSTn}$ after fitting RF classification function &
    some \\
\hline
\end{tabular}
}
\end{center}
\end{table}
\setlength{\extrarowheight}{0pt}

The five estimation methods we used are listed in Table \ref{tab:sims_methods}. \lq\lq Raw Comparison" means only empirical averages were compared by taking their difference; i.e., no modeling was performed. We expected each GLM-based estimator to be unbiased because it correctly specified the relevant data-generating mechanism (i.e., both the model form and covariate set). In comparison, each RF-based estimator could have been more biased at a given sample size because it only correctly specified the covariates of the relevant data-generating mechanism (i.e., not the model form). However, for the simple linear data-generating mechanisms we used, we nonetheless expected both RF methods to have been more statistically consistent (i.e., asymptotically less biased as $m$ grows large) than the raw comparison estimator.

To implement MoTR, we used between $r=10$ and $r=200$ MoTR runs, and applied trimming (i.e., removing extreme values at the 5th and 95th percentiles) and overlapping to our stabilized propensity scores. (See the Appendix for details.) We also followed a stopping rule based on the degree of convergence over MoTR runs measured using a modified coefficient of variation (see Appendix).

The MoTR-GLM and PSTn-GLM methods all fit the correct models (i.e., the correct ARCO outcome or exposure mechanisms specified in the Appendix for Subsection \ref{subsec:sims_dgp}). To implement the RF methods, we fit prediction functions. For the MoTR-RF method, we fit the prediction function $Y_t \sim f \big( X_t, Y_{t-1} \big)$. For the PSTn-RF method, we fit the classification function $X_t \sim f \big( Y_{t-1} \big)$.

\setlength{\extrarowheight}{4pt}
\begin{table}
\caption{ \label{tab:sims_results_one} Simulation study analysis results for one simulated dataset with a true APTE of 1.89.}
\begin{center}
\scalebox{1}{
\begin{tabular}{ l | l | r }
\hline
Method & Estimated APTE (95\% CI) & Bias \\
\hline
Raw Comparison & $1.76 \; (1.59, 1.93)$ & $-0.13$ \\
MoTR-GLM & $1.86 \; (1.69, 2.04)$ & $-0.03$ \\
MoTR-RF & $1.84 \; (1.62, 2.07)$ & $-0.05$ \\
PSTn-GLM & $1.94$ & $0.05$ \\
PSTn-RF & $1.88$ & $-0.01$ \\
\hline
\end{tabular}
}
\end{center}
\end{table}
\setlength{\extrarowheight}{0pt}

\clearpage
\setlength{\extrarowheight}{4pt}
\begin{table}
\caption{ \label{tab:sims_results_all_1461} Simulation study analysis results for 100 simulated datasets ($m = 1461$ each) with a true APTE of 1.89.}
\begin{center}
\scalebox{1}{
\begin{tabular}{ l | l }
\hline
Method & Estimated Mean Bias (95\% CI) \\
\hline
Raw Comparison & $-0.15 \; (-0.17, -0.14)$ \\
MoTR-GLM & $-0.01 \; (-0.02, \; 0.00)$ \\
MoTR-RF & $-0.10 \; (-0.11, -0.09)$ \\
PSTn-GLM & $-0.01 \; (-0.04, \; 0.02)$ \\
PSTn-RF & $-0.00 \; (-0.04, 0.03)$ \\
\hline
\end{tabular}
}
\end{center}
\end{table}
\setlength{\extrarowheight}{0pt}

For each method, the APTE estimated from one simulated dataset is shown in Table \ref{tab:sims_results_one}. MoTR and PSTn estimate the true APTE with little bias compared to the raw comparison when their models are specified correctly (i.e., MoTR-GLM and PSTn-GLM, respectively). Both MoTR-RF and PSTn-RF use the correct predictors, but not the true models. This results in lower statistical efficiency; i.e., at a given sample size, the misspecified RF estimator can be more biased or vary more widely than its GLM counterpart. And indeed, the MoTR-RF CI (size = 0.45) is wider than the MoTR-GLM CI (size = 0.35).

\begin{figure} 
    \centering
    \includegraphics[width=1\linewidth]{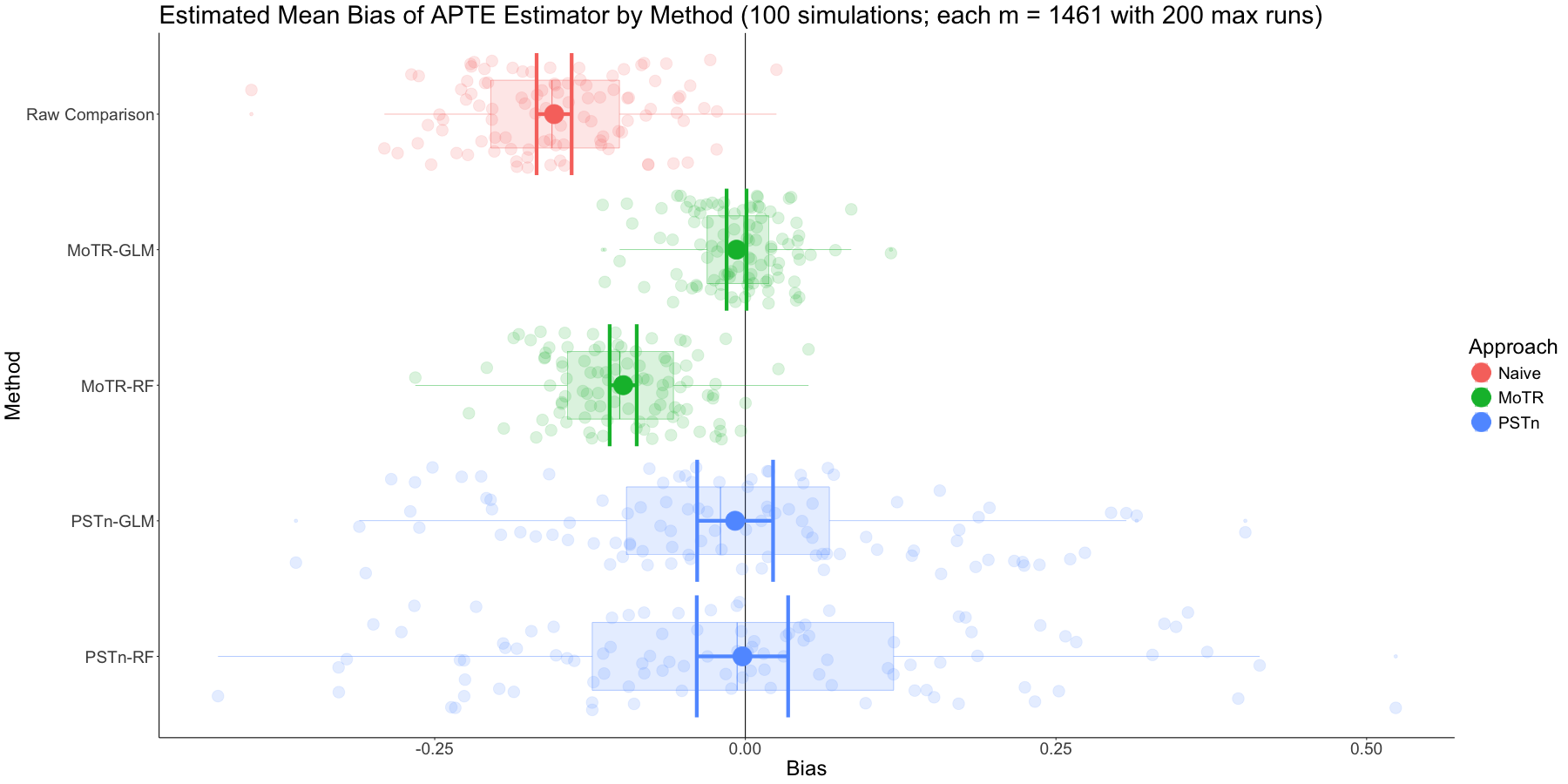}
    \caption{Simulation study analysis results for 100 simulated datasets ($m = 1461$ each) with a true APTE of 1.89. Each small light dot represents the bias of one simulated dataset for estimating a true APTE of 1.89. Each big dark dot represents the average bias (i.e., estimated mean bias) over all 100 datasets, with corresponding 95\% confidence interval shown as symmetric error bars.}
    \label{fig:sims_100_1461}
\end{figure}

For each method, the estimated mean bias (i.e., approximate accuracy) of the estimator at $m = 1461$ is summarized over 100 simulated datasets in Table \ref{tab:sims_results_all_1461} and Figure \ref{fig:sims_100_1461}. The estimated mean bias for the raw comparison method is the largest compared to all others. Over many simulated datasets, it is clear that both PSTn estimators, while fairly unbiased, vary more than both MoTR estimators. Over about 1,000 to 1,100 MoTR runs of $n=1461$ each (i.e., 100 simulations, each with about 10-11 MoTR runs until convergence), the coverage of the MoTR-GLM method was 94.1\%, lower than the nominal 95\% coverage probability. This was as expected; see the end of Subsection \ref{subsec:motr-procedure}.

\section{Empirical Study}\label{sec:empirical}

Co-authors EJD and LS comprised our two study participants. In this Section, we describe how we estimated an APTE of daytime walking cadence on TST that night. We did this separately for each participant. In our models, we included an indicator for weekend (versus weekday), which we treated as an exogenous variable. To reflect the fact that we would not be able to modify the day of the week in a hypothetical n-of-1 experiment, we did not modify its original values. Hence, our estimand was that of \eqref{eqn:apte_explicit}---a partially transportable APTE.

Similar to the simulation study of Section \ref{sec:sims}, we analyzed $m=1374$ days of EJD's sleep duration and step-count data using a Fitbit Charge\texttrademark{} wrist-worn sensor, collected from June 2020 through March 2024. EJD wore a Fitbit Charge\texttrademark{} 3, followed by a Fitbit Charge\texttrademark{} 4, and finally a Fitbit Charge\texttrademark{} 6. We dichotomized our exposure based on the observed median of 3.8 steps per minute over all 1,374 days of observation.

We also analyzed $m=2800$ days (about 7 years and 8 months) of LS's sleep duration and step-count data using a Fitbit Charge\texttrademark{} wrist-worn sensor, collected from April 2015 through July 2024. LS wore a Fitbit Charge\texttrademark{} 2, followed by a Fitbit Inspire\texttrademark{} 2, and finally a Fitbit Sense\texttrademark{} smartwatch. For LS, the observed median walking cadence over all 2,800 days of observation was 5.6 steps per minute. Before fitting any models, we squared the sleep duration outcome in order to better meet the normality assumption required by the t-tests used.

We used the same five estimation methods listed in Table \ref{tab:sims_methods}. As with our simulation study, we used between $r=10$ and $r=200$ MoTR runs with the same convergence-based stopping rule, and applied trimming and overlapping to our stabilized propensity scores (see Appendix). For EJD, propensities were trimmed at the 5th and 95th percentiles. LS had extremely small or large propensities, so these were trimmed at the 20th and 80th percentiles. For all four methods, $\bm{W}_t$ consisted of the endogenous variables $X_{t-1}$ and $Y_{t-1}$,
and the exogenous variable $W^{ex}_t$ where $W^{ex}_t = 1$ if day $t$ was a weekend, and $W^{ex}_t = 0$ otherwise.

We fit the following outcome models. MoTR-GLM fit the ARCO model $Y_t = \beta_0 + \beta_X X_t + \beta_{co} X_{t-1} + \beta_{ar} Y_{t-1} + \beta_{Xar} X_t Y_{t-1} + \beta_{ex} W^{ex}_t + \calE_t$. MoTR-RF fit the prediction function $Y_t \sim f \big( X_t, X_{t-1}, Y_{t-1}, W^{ex}_t \big)$. Note that we did not include the interaction term $X_t Y_{t-1}$ because RF automatically segments or partitions the values of $X_t$ and $Y_{t-1}$ in a way that allows for interactions. MoTR used $r_{min} = 10$ to $r_{max} \le 200$ runs. We used the same stopping rule as in Subsection \ref{subsec:sims_results}.

We fit the following propensity models. PSTn-GLM fit the model $\logit \big( \pi_t \big) = \alpha_0 + \alpha_{ar} X_{t-1} + \alpha_{en} Y_{t-1} + \alpha_{ex} W^{ex}_t$, where $\alpha_{ar}$ denotes autoregression with respect to $X$ (rather than $Y$), and $\pi_t = \Pr \big( X_t = 1 | X_{t-1}, Y_{t-1}, W^{ex}_t \big)$. PSTn-RF fit the classification function $X_t \sim f \big( X_{t-1}, Y_{t-1}, W^{ex}_t \big)$.

\setlength{\extrarowheight}{4pt}
\begin{table}
\caption{ \label{tab:real_results} Empirical study analysis results: Estimated APTEs and 95\% CIs for authors EJD and LS. APTE specified as mean hours of sleep after a day of walking quickly (i.e., $>$ median cadence) minus that after a day of walking slowly (i.e., $\le$ median cadence).}
\begin{center}
\scalebox{1}{
\begin{tabular}{ l | l | l }
\hline
Method & EJD & LS \\
\hline
Raw Comparison & $0.55 \; (0.36, 0.74)$ & $-1.76 \; (-2.05, -1.43)$ \\
MoTR-GLM & $0.46 \; (0.27, 0.65)$ & $-1.61 \; (-1.91, -1.23)$ \\
MoTR-RF & $0.27 \; (0.08, 0.47)$ & $-1.29 \; (-1.66, -0.77)$ \\
PSTn-GLM & $0.37$ & $-0.65$ \\
PSTn-RF & $-0.48$ & $-2.94$ \\
\hline
\end{tabular}
}
\end{center}
\end{table}
\setlength{\extrarowheight}{0pt}

The estimated APTEs for both EJD and LS are shown in Table \ref{tab:real_results}. For LS, the estimates and CIs (where applicable) have been transformed back to the original scale by taking the square root of the original values; hence the asymmetric CIs.

After a fast-cadence day, EJD slept 0.55 hour (i.e., just over 30 minutes) more on average than he did after a slow-cadence day. However, if his MoTR-GLM model reflected reality (by representing the true outcome mechanism), then walking quickly in fact only caused EJD to sleep 0.46 hour more that night on average. If instead his MoTR-RF model reflected reality, then walking quickly caused EJD to sleep only 0.27 hour (i.e., about 15 minutes) more that night on average. In either case, 0.55 hour would have been a naive overestimate of how much additional sleep EJD got on average as a result of walking quickly versus slowly.

If EJD's PSTn-GLM model reflected reality (by representing the true propensity mechanism), then walking quickly caused EJD to sleep only 0.37 hour (just over 20 minutes) more that night on average. This qualitatively resembled both MoTR findings. However, if his PSTn-RF model reflected reality, then walking quickly caused EJD to sleep 0.48 hour less that night on average---the opposite in both sign and magnitude from the other findings. In the latter case, raw comparison would have led EJD to mistakenly conclude that walking quickly caused him to sleep about half an hour longer on average, when in fact it made him sleep about half an hour less.

Contrast EJD's findings to those of LS. After a fast-cadence day, LS got 1.76 fewer hours of sleep on average than he did after a slow-cadence day. However, if his MoTR-GLM model reflected reality, then walking quickly in fact only caused LS to sleep 1.61 hours less that night on average. If instead his MoTR-RF model reflected reality, then walking quickly caused LS to sleep only 1.29 hours less that night on average. In either case, 1.76 hours would have been a naive overestimate of how much less sleep LS got on average as a result of walking quickly versus slowly.

If LS's PSTn-GLM model reflected reality, then walking quickly in fact only caused LS to sleep 0.65 hours (about 40 minutes) less that night on average---at most half of the effects estimated using either MoTR model. If instead his PSTn-RF model reflected reality, then walking quickly caused LS to lose more sleep (2.94 hours that night on average) than any of the other models. In the latter case, raw comparison would have led LS to mistakenly conclude that walking quickly caused him to lose just under two hours of sleep on average, when in fact it made him lose almost three hours of sleep.

\section{Discussion}\label{sec:discuss}

\subsection{Summary of Findings}\label{subsec:summary-of-findings}

We presented a new method, model-twin randomization or MoTR, to estimate the average period treatment effect using wearable sensor data. We believe that MoTR is an intuitive approach that statisticians, computer scientists, clinicians, and patients can readily understand, collaborate on, and implement.

We used MoTR to estimate the daily APTE of walking cadence on sleep duration using 1,374 and 2,800 days of EJD's and LS's Fitbit\texttrademark{} data, respectively. The striking differences in estimated APTEs between EJD and LS illustrate the need for using flexible n-of-1 or time series methods over more rigid approaches like hierarchical models.

Our simulation results confirmed the trends in bias we expected, listed in Table \ref{tab:sims_methods}. Interestingly, while the MoTR-RF estimator was less biased than the raw comparison estimator at $m = 1461$, it was more biased than the other three estimators. We theorized that the MoTR-RF estimator is still more statistically consistent than the raw comparison estimator.

To test this theory, we ran our simulation study using a larger sample size of eight years (i.e., $m = 2922$ days) of data. The results are shown in Appendix Table
6
and Appendix Figure
A3.
With the larger sample size, all CIs were smaller (as expected). And the estimated mean bias for the MoTR-RF estimator in particular was smaller than that observed at the smaller sample size of $m = 1461$---lending some support to our theory about its asymptotic behavior.

For each study participant, the variability in APTE estimates across all four methods (e.g., in opposite direction or order of magnitude) highlights the importance of specifying a model that resembles the true mechanism as much as possible. This should be done both empirically (i.e., in a data-driven way; for example, through parameter tuning and model selection) and scientifically (i.e., based on related literature). Our findings provide a real example of why researchers sought to develop the doubly robust methods mentioned in Subsection \ref{subsec:gfstats}.

\subsection{Limitations, Extensions, and Future Directions}\label{subsec:extensions-and-future-directions}

At least two key limitations must be surmounted before MoTR can truly find practical use. Real-world settings will likely involve longer time dependencies, as well as multivariate time series of more than two variables. The theory we developed is complex enough that we chose to limit our application of MoTR to the case of one lag and two time series. We hope that readers will find this simple case helpful in understanding the concepts developed. Future work should be done to better characterize MoTR's performance under more realistic settings involving more lags and richer multivariate time series data. In particular, MoTR performance when modeling longer time dependencies might be improved through variance reduction techniques.

MoTR provides a way to surface plausible or suggested causal effects for possible further investigation or confirmation through an n-of-1 experiment. It can be used to develop a personalized idiographic intervention plan, as done by Yoon et al (2018)\cite{2018-yoon-etal}. For example, one might use a data-driven procedure like that shown in the Appendix to identify combinations of models and plausible confounders that produce APTE estimates that meaningfully differ from naive estimates calculated by comparing raw averages (as in Table \ref{tab:sims_methods}).

We are happy to report that work is already underway to deepen idiographic causal approaches in both theoretical and applied directions. For a recent survey of causal inference methods for time series data, see Moraffah et al (2021) \cite{2021-moraffah-etal}.

An advanced doubly robust approach is being developed by Malenica et al (2021)\cite{2021-malenica-etal}. They use targeted maximum likelihood estimation (TMLE) to estimate and infer the APTE. This important paper lays the theoretical groundwork for both the asymptotic consistency and normality of their TMLE estimator, and proposes a sequentially adaptive design for learning the optimal idiographic treatment rule over time. The TMLE approach also optimizes the use of machine learning models by minimizing the bias incurred when fitting such models. For example, Matias and Wac (2022)\cite{2022-matias-wac} applied a deep learning method (recurrent neural networks) for time series prediction. Malenica et al (2023) \cite{2023-malenica-etal} have since developed \lq\lq a promising [design-based] framework for [anytime-valid] statistical inference ... at all time points of an N-of-1 trial" that extends the PSTn theory sketched out in Subsection \ref{subsec:propensity-score}.

Intraday sensor data can also be used to characterize average trends---functionals, not just scalars---across periods, analogous to longitudinal trends across individuals. In a nod to the spaghetti plots of longitudinal trend analysis, Daza (2019) \cite{2019-daza} introduced the term {\sl pancit plot} to describe a plot of intra-period trends. (\lq\lq Pancit" is pronounced like \lq\lq pun-SEAT"; it is Tagalog/Filipino for \lq\lq noodles".) Daza et al (2020) \cite{2019_daza_etal} illustrated pancit plots for continuous blood glucose intraday data, in which the APTE was not a scalar difference, but a difference in a daily functional trends; i.e., the difference in slope and time polynomial coefficients of a mixed-effects model to describe both the within-period trend and variability of repeated measures under each treatment condition. The clear implication is to extend APTE-based techniques (including MoTR and PSTn) to functional data analysis \cite{2015-goldsmith-etal} of intensive longitudinal data in order to estimate other types of APTEs and RITEs.

The APTE framework was created to enable and facilitate this extension. Each time point $t(j)$ can contain sub-points (e.g., $t(j_k)$). This modularity allows for flexible temporal scaling of posited causal relationships. The framework's utility can also be complemented and improved with formal causal diagrams like DAGs for conceptualizing formal structures.

Finally, to compare oneself to others, one might ask, how does my APTE relate to a corresponding group-level (nomothetic) ATE? One common approach is to combine APTE estimates in the same way ATEs from multiple nomothetic studies would be aggregated in a meta-analysis (i.e., a {\sl series-of-n-of-1}). Another approach is to use a mixed-effects model to combine and compare participant-level APTEs to the overall ATE; see the Appendix Section on hierarchical models. Zucker et al (1997, 2010)\cite{1997_zucker_etal, 2010_zucker_etal} examine these and other approaches in detail. Their work helps build a foundation with which to extend our single-subject approach for functional data analysis.

More generally, a number of researchers have characterized when an n-of-1 approach can be more useful than a standard group-based approach (e.g., \cite{2016-beltz-etal, 2016-araujo-etal, 2019-blackston-etal}). This includes recent work by Golino et al \cite{2022-golino-etal} to define a metric called the \lq\lq ergodicity information index" to help guide such decisions on study design and analysis.

\subsection{It's About Time}\label{subsec:its-about-time}

Within-individual approaches are increasing in popularity thanks to the growing availability of small data and idiographic measurements. Most recently, these include \lq\lq genetically individualized n-of-1 trials" for treating people with rare conditions and diseases \cite{2015_schork, 2024-augustine-etal}.

This larger field of quantitative idiographic approaches beyond just n-of-1 and single-case designs is what author EJD calls \lq\lq esametry" \cite{2024-daza}, derived from \lq\lq isa" (pronounced \lq\lq ee-SA"), the Tagalog Filipino word for \lq\lq one". Econometrics, psychometrics, and biostatistics are the respective applications of statistics to economics, psychological measurement, and clinical trials and population health. Likewise, esametry (or esametrics) is the application of statistics to a single person, individual, or unit.

Statistics seeks to understand and shape populations. Esametry seeks to understand and shape ourselves. History now calls on statisticians and data scientists to apply their formidable quantitative and computational skills to the rigorous esametric analysis of such personally meaningful data. It's about time.

\section{Acknowledgements}

This paper was fully supported by EJD's \href{https://statsof1.org/post/}{newsletter} and \href{https://creators.spotify.com/pod/profile/n-of-5-minutes/}{podcast} (\href{https://statsof1.org/}{Stats-of-1}) while he was a full-time employee at Evidation, by IM's current institution (University of Geneva), and by LS's current employers (Stanford Medicine, Alphabet). Early formative work on this paper was supported by the Stanford Clinical and Translational Science Award (CTSA) to Spectrum (UL1 TR001085). The CTSA program is led by the National Center for Advancing Translational Sciences (NCATS) at the National Institutes of Health (NIH). This research was also supported by NIH grant 2T32HL007034-41 that funded EJD's postdoctoral training. IM was supported by AGE-INT Swissuniversities, Centre Universitaire d’Informatique of the University of Geneva, and Société Académique de Genève. LS has no relevant grant funding to report. The content is solely the responsibility of the authors and does not necessarily represent the official views of the NIH, other funding agencies, or the authors' employers.

EJD thanks the following people. Igor Matias for his constant feedback in ideation, which inspired EJD to come up with the acronym \lq\lq MoTR". Professor Linda Valeri for her helpful input on theoretical considerations. Professor Jared Huling for his direction in understanding the initially unexpected random forests results. Professor Michael Baiocchi for his constant support, feedback, and guidance.

We thank our manuscript referees for helping us significantly improve the paper, particularly through the refinement or addition of material in Sections and Subsections \ref{subsec:tempconsids}, \ref{subsec:deriving-the-apte}, \ref{subsec:modflex}, \ref{subsec:idiographic-approaches}, \ref{subsec:motr}, \ref{sec:sims}, and \ref{subsec:extensions-and-future-directions}. We also thank our family and friends for their constant support; our colleagues, teachers, and students; and of course, the luck and privilege that gave us this opportunity to improve personalized health and medicine.

EJD dedicates this paper to Filipinos and Filipino-Americans. To all who are underrepresented or unacknowledged in science, technology, engineering, and math (STEM), and in academia: Kaya natin 'to! To you, the reader: Know yourself, help others, and find meaning in all things.

\bibliographystyle{vancouver}
\bibliography{references}

\bigskip
\noindent
A comprehensive list of references on statistical methods for within-individual studies can be found at \url{https://statsof1.org/resources/}.

\clearpage
\title{\bfseries Appendix for \lq\lq Model-Twin Randomization (MoTR) for Estimating the Recurring Individual Treatment Effect"}
\author{\bfseries Eric J. Daza, DrPH, MPS; Igor Matias, MSc; Logan Schneider, MD}

\maketitle

\setcounter{figure}{0}

\makeatletter 
\renewcommand{\thefigure}{A\@arabic\c@figure}
\makeatother

\noindent
The following material is organized by relevant Section and Subsection numbers found in the main manuscript, but prefixed with \lq\lq A".

\section*{A1 Introduction}

\subsection*{A1.1 Within-Individual Studies}

\subsubsection*{A1.1.1 Relationship to Hierarchical Models, Bayesian Inference, and Data Table Pivots}

Analytically, the paradigm shift from hierarchical to within-individual studies is made precise at the level of inference and in the population's temporal properties. The key estimand of a hierarchical study is a {\sl group-level average} across a {\sl contemporaneous} population, where autocorrelation induced by repeated measurements is a \lq\lq nuisance" to conducting inference properly (often involving nuisance parameters). The key estimand of a within-individual study is an {\sl individual-level average} across a {\sl sequential} population; the nature of any autocorrelation may sometimes itself be of interest (e.g., as an effect modifier or moderator).

Consider the motivation behind the mixed-effects model. A \lq\lq complete pooling" approach assumes that outcomes can systematically vary by groups of individuals (via \lq\lq fixed effects" terms), but do not also systematically vary by individual. This model would not include any terms for identifying individuals. A \lq\lq partial pooling" approach assumes that outcomes can also systematically vary by individual (via \lq\lq random effects" terms). This assumption is operationalized by the standard mixed-effects model. A \lq\lq no pooling" approach seeks to model outcomes separately for each individual, without trying to explain systematic variation in outcomes over groups of individuals. The latter is the n-of-1 approach.

In data-engineering terms, consider the following data table. Each column denotes a variable, and each row denotes sequentially measured data values per individual. Rows are grouped by individuals uniquely identified by an identifier column. To shift the analytic focus from between-individual group-level inference to separate per-individual inference, one could first pivot this table from \lq\lq long to wide".

For example, pivot the table, grouped by identifier, such that every row comprises exactly one individual's full n-of-1 dataset. The latter would contain groups of columns, each group representing a particular variable measured repeatedly over time. Each row would be an \lq\lq n-of-1, m-of-many", where $m$ denotes the number of time points, phases, or periods of repeated measurements.

A Bayesian approach might involve using one's personal beliefs and history to elicit the prior distribution of person-specific ARCO parameters. The epistemological focus of a within-individual study is on one person---at a time, at least. Hence, it is reasonable to primarily (if not wholly) rely on that particular individual's experiences in constructing a priori hypotheses when feasible. Consideration or analysis of previously collected self-tracked (i.e., collected on oneself) data can help refine these hypotheses, as can group-level scientific findings (e.g.., the latter might be used to set starting values for computationally iterative approaches).

\section*{A2 Methodological Theory}

\subsection*{A2.2 Causal Inference}

\subsubsection*{A2.2.1 Core Concepts}

\noindent
\textbf{A2.2.1.1 Why Use Potential Outcomes?}

\smallskip
The PO framework is generally not needed to conduct causal inference in simple or straightforward cases. From our example, we could just evaluate the model $Y = \beta_0 + \beta_1 X + \bm{W} \bm{\beta}_2 + \calE$ directly when $X$ is randomized. So why use it?

The power of the PO framework, elaborated in Section \ref{subsec:ate}, is in handling the complexity of commonplace cases that have become particularly ubiquitous in the era of data science. These span both observational studies with their non-randomized predictors (e.g., convenience samples, real-world data, health claims, medical records, user behavior) and experimental studies with their randomized interventions (e.g., A/B testing, experimentation, network effects, etc.).

Thinking in terms of POs also implicates methods of statistical inference that are far from obvious when one is mainly concerned with how to statistically predict or model an outcome---the main focus of modern data science and machine learning. For example, the PO framework allows causal inference to be understood as a missing data or survey sampling problem, or to be conducted via randomization-based inference (in contrast to the superpopulation-based inference central to statistical modeling and machine learning).

This key insight is the reason we now have propensity score matching and inverse probability weighting (IPW) methods \cite{1983_rosenbaum_rubin}. POs also encourage development, implementation, and detailed characterization and analysis of discrete interventions that are intuitive and actionable (e.g., \lq\lq do A=1 or A=0").

\smallskip
\noindent
\textbf{A2.2.1.2 Directed Acyclic Graph of Core Assumptions}

\begin{figure}

	\centering

	\begin{tikzpicture}

		\node(Y1)[]{$Y^1$};
		\node(W)[left = of Y1]{$\bm{W}$};
		\node(X)[above = of Y1]{$X$};
		\node(Y0)[below = of Y1]{$Y^0$};
		\node(Y)[right = of Y1]{$Y$};



		\draw [->] (W) -- (Y1);

		\draw [->] (W) -- (Y0);

		\draw [->] (X) -- (Y);
		\draw [->] (Y1) -- (Y);
		\draw [->] (Y0) -- (Y);

	\end{tikzpicture}

	\caption{Directed acyclic graph (DAG) for a randomized binary exposure.}\label{fig:dagXrandomized}

\end{figure}

\begin{figure}

	\centering

	\begin{tikzpicture}

		\node(Y1)[]{$Y^1$};
		\node(W)[left = of Y1]{$\bm{W}$};
		\node(X)[above = of Y1]{$X$};
		\node(Y0)[below = of Y1]{$Y^0$};
		\node(Y)[right = of Y1]{$Y$};


		\draw [->] (W) -- (X);

		\draw [->] (W) -- (Y1);

		\draw [->] (W) -- (Y0);

		\draw [->] (X) -- (Y);
		\draw [->] (Y1) -- (Y);
		\draw [->] (Y0) -- (Y);

	\end{tikzpicture}

	\caption{Directed acyclic graph (DAG) for a non-randomized binary exposure.}\label{fig:dagXnotrandomized}

\end{figure}

\smallskip
Figure \ref{fig:dagXrandomized} illustrates the relationships between the observed outcome $Y$, binary intervention $X = s$ where $s \in \{ 0, 1 \}$, potential outcomes $\{ Y^0, Y^1 \}$, and $\bm{W}$ (i.e., the set of all other direct or indirect causes of $Y$ that may also be direct or indirect causes of $X$, or may contextualize the effect of $X$ on $Y$) when $X$ is randomized. Figure \ref{fig:dagXnotrandomized} illustrates the case when $X$ is not randomized.

	\subsubsection*{A2.5.2 APTE Derivation}

The general formula for the CAPO resembles the causal consistency formula:
\begin{align*}
    Y^{s_t \bigcdot}_t
        &=
        	E_{\overline{\bm{X}}_t} \left(
        	    Y^{s_t \overline{\bm{X}}_t}_t \big| X_t = s_t
            \right) \nonumber \\
        &=
        	E_{\overline{\bm{X}}_t} \left\{
            	\sum_{\{ \overline{\bm{s}}_t \}}
            	    Y^{s_t \overline{\bm{s}}_t}_t I \left( \overline{\bm{X}}_t = \overline{\bm{s}}_t \right)
        	    \Bigg| X_t = s_t
            \right\} \text{ using the serial causal consistency formula} \nonumber \\
        &=
        	\sum_{\{ \overline{\bm{s}}_t \}}
        	    Y^{s_t \overline{\bm{s}}_t}_t 
            	E_{\overline{\bm{X}}_t} \left\{
                	 I \left( \overline{\bm{X}}_t = \overline{\bm{s}}_t \right)
            	    \Big| X_t = s_t
                \right\} \nonumber \\
        &=
        	\sum_{\{ \overline{\bm{s}}_t \}}
        	    Y^{s_t \overline{\bm{s}}_t}_t
        	    \sum_{\{ \overline{\bm{x}}_t \}}
                    I \left( \overline{\bm{x}}_t = \overline{\bm{s}}_t \right)
        	        \Pr \left( \overline{\bm{X}}_t = \overline{\bm{x}}_t \big| X_t = s_t \right) \nonumber \\
        &=
        	\sum_{\{ \overline{\bm{s}}_t \}}
        	    Y^{s_t \overline{\bm{s}}_t}_t
        	    \Pr \left( \overline{\bm{X}}_t = \overline{\bm{s}}_t \big| X_t = s_t \right)
\end{align*}
The APTE for an observation period of length $m$, $\delta^\text{APTE}_{(m)}$ is equal to the mean HAPTE over this same observation period, $\delta^\text{HAPTE}_{(m)}$, taken over all possible randomized exposure histories:
\begin{align}
    \delta^\text{APTE}_{(m)}
        &=
            E^{(m)} \left(
                \delta^\text{PTE}_t
            \right)
            \nonumber \\
        &=
            E^{(m)} \left\{
                \delta^\text{PTE}_t \left( \overline{\bm{X}}_t \right)
            \right\}
            \nonumber \\
        &=
            \frac{1}{m} \sum_{t=1}^m
            E \left\{
                \delta^\text{PTE}_t \left( \overline{\bm{X}}_t \right)
            \right\}
            \nonumber \\
        &=
            \frac{1}{m} \sum_{t=1}^m
            E \left\{
                \delta_t \left( \overline{\bm{X}}_t \right)
            \right\}
            \text{ by historical PTE definition} \nonumber \\
        &=
            \frac{1}{m} \sum_{t=1}^m
            E_{\overline{\bm{X}}_t, \calE_t} \left\{
                \delta_t \left( \overline{\bm{X}}_t \right)
            \right\}
            \nonumber \\
        &=
            \frac{1}{m} \sum_{t=1}^m
            E_{\calE_t} \left[
                E_{\overline{\bm{X}}_t} \left(
                    \delta_t \left( \overline{\bm{X}}_t \right)
                    \big| \calE_t
                \right)
            \right]
            \nonumber \\
        &=
            \frac{1}{m} \sum_{t=1}^m
            E_{\calE_t} \left[
                E_{\overline{\bm{X}}_t} \left(
                    \delta_t \left( \overline{\bm{X}}_t \right)
                \right)
            \right]
            \nonumber \\
        &=
            \frac{1}{m} \sum_{t=1}^m
            E_{\calE_t} \left\{
                \delta_t ( \bigcdot )
            \right\}
            \nonumber \\
        &=
            E \left\{
                \delta_t ( \bigcdot )
            \right\}
            \nonumber \\
        &=
            \delta^\text{HAPTE}_{(m)}
        \label{eqn:apte_appendix}
\end{align}

To see how this APTE can be estimated using only observed data (i.e., without knowing any counterfactuals), recall that the conditional mean observed outcome at a given period when $X$ is randomized at every period is equal to the mean CAPO at $t$:
\begin{align}
    E( Y_t | X_t = s_t )
        &=
            E_{\calE_t} \left\{
                E \left(
                    Y_t
                    | X_t = s_t, \calE_t
                \right)
                | X_t = s_t
            \right\}
            \nonumber \\
        &=
            E_{\calE_t} \left\{
                E \left(
                    Y_t
                    | X_t = s_t, \calE_t
                \right)
            \right\}
            \text{ because } \calE_t \indep X_t
            \nonumber \\
        &=
            E_{\calE_t} \left\{
                E_{\overline{\bm{X}}_t} \left(
                    Y^{X_t \overline{\bm{X}}_t}_t
                    | X_t = s_t, \calE_t
                \right)
            \right\}
            \text{ by serial causal consistency} \nonumber \\
        &=
            E_{\calE_t} \left\{
                E_{\overline{\bm{X}}_t} \left(
                    Y^{s_t \overline{\bm{X}}_t}_t
                    | X_t = s_t, \calE_t
                \right)
            \right\}
            \nonumber \\
        &=
            E_{\calE_t} \left\{
                E_{\overline{\bm{X}}_t} \left(
                    Y^{s_t \overline{\bm{X}}_t}_t
                    | \calE_t
                \right)
            \right\}
            \text{ by randomization of } X_t
            \nonumber \\
        &=
            E_{\calE_t} \left\{
                E_{\overline{\bm{X}}_t} \left(
                    Y^{s_t \overline{\bm{X}}_t}_t
                \right)
            \right\}
            \text{ because } \calE_t \indep \overline{\bm{X}}_t
            \nonumber \\
        &=
            E ( Y^{s_t \bigcdot}_t )
        \label{eqn:eytbarxt}
\end{align}

\section*{A3 Estimation Methods}

	\subsection*{A3.2 Idiographic Approaches}

The general PSTn procedure is:
\begin{enumerate}

    \item Fit the propensity model $\pi_t = \Pr \big( X_t = 1 | \bm{W}_t \big)$ to the data. Similar to MoTR, the estimator $\hat{\pi}_t$ from fitting the model is the participant's {\sl propensity score twin}---a digital twin that represents the participant's exposure mechanism for estimating the APTE.
    
    \item Weight each observed $Y_t$ by the reciprocal of its corresponding estimated propensity based on the observed exposure $x_t$ as follows. This reciprocal is commonly called an \lq\lq inverse probability weight" or \lq\lq inverse propensity weight".
    
    \begin{enumerate}
    
        \item Calculate $\widehat{ipw}_t = 1 \big/ \big\{ x_t \hat{\pi}_t + \big( 1 - x_t \big) \big( 1 - \hat{\pi}_t \big) \big\}$.
        
        \item Consider applying the following adjustments to mitigate overly large weights resulting from overly small estimated propensities.
        
        \begin{enumerate}
        
            \item Only use trimmed $\hat{\pi}_t$ values (i.e., drop extreme $\hat{\pi}_t$ values). For example, drop all $\hat{\pi}_t$ values (and their corresponding $Y_t$ values) less than the 5th percentile or greater than the 95th percentile of all $\hat{\pi}_t$ values. 
            
            \item Only use overlapping $\hat{\pi}_t$ values (i.e., keep the range of $\hat{\pi}_t$ values with the most overlap between groups with $s \in \{ 0, 1 \}$). For example, first calculate the range (i.e., minimum and maximum values) of $\widehat{ipw}_t$ values with corresponding observed exposure $x_t = 1$. Then calculate the range of $\widehat{ipw}_t$ values with corresponding observed exposure $x_t = 0$. Finally, calculate the overlapping range as follows: the minimum is the maximum of both minimum values, and the maximum is the minimum of both maximum values. Only keep $\hat{\pi}_t$ values (and their corresponding $Y_t$ values) within the overlapping range.
            
            \item Use stabilized weights (i.e., multiply $\Tilde{Y}_t$ by the proportion of the sample with $X = s_t$ for the corresponding value of $x_t$) \cite{2000_robins_etal}.
            
        \end{enumerate}

        \item Multiply $Y_t$ by its IPW as $\Tilde{Y}_t = Y_t \widehat{ipw}_t$.
            
    \end{enumerate}
    
    \item For $s \in \{ 0, 1 \}$, estimate $E \big( Y^s \big)$ as the average of the weighted outcomes; i.e., $\bar{\Tilde{Y}}^s = \frac{1}{m_s} \sum_{t=1}^m \Tilde{Y}_t I \big( X_t = s \big)$.
    
    \item Estimate the APTE as $\hat{\delta}^\text{PSTn} = \bar{\Tilde{Y}}^1 - \bar{\Tilde{Y}}^0$.

\end{enumerate}

    	\subsubsection*{A3.2.1 Idiographic G-Formula and APTE Transportability}

\noindent
The mean CAPO for an observation period of length $m$ is derived as follows:
{\small
\begin{align*}
    & E \left(
        Y_t
        \big| X_t = s_t, \overline{\bm{R}}_t
    \right)
    \nonumber \\
        &=
            E \left(
                Y^{X_t \overline{\bm{X}}_t}_t \Big| X_t = s_t, \overline{\bm{R}}_t
            \right) \text{ by serial causal consistency} \nonumber \\
        &=
            E \left(
                Y^{s_t \overline{\bm{X}}_t}_t \Big| X_t = s_t, \overline{\bm{R}}_t
            \right) \nonumber \\
        &=
            E^{(m)}_{\bm{W}^{en}_t, \calE_t, \bm{W}^{ex}_t} \left(
                Y^{s_t \overline{\bm{X}}_t}_t
                \Big| X_t = s_t, \overline{\bm{R}}_t
            \right)
            \nonumber \\
        &=
            E^{(m)}_{\overline{\bm{X}}_t, \calE_t, \bm{W}^{ex}_t} \left(
                Y^{s_t \overline{\bm{X}}_t}_t
                \Big| X_t = s_t, \overline{\bm{R}}_t
            \right)
            \nonumber \\
        &=
            E^{(m)}_{\bm{W}^{ex}_t} \left\{
                E_{\overline{\bm{X}}_t, \calE_t} \left(
                    Y^{s_t \overline{\bm{X}}_t}_t
                    \Big| \bm{W}^{ex}_t, X_t = s_t, \overline{\bm{R}}_t
                \right)
                \Big| X_t = s_t, \overline{\bm{R}}_t
            \right\}
            \nonumber \\
        &=
            E^{(m)}_{\bm{W}^{ex}_t} \left[
                E_{\calE_t} \left\{
                    E_{\overline{\bm{X}}_t} \left(
                        Y^{s_t \overline{\bm{X}}_t}_t
                        \Big| \calE_t, \bm{W}^{ex}_t, X_t = s_t, \overline{\bm{R}}_t
                    \right)
                    \Big| \bm{W}^{ex}_t, X_t = s_t, \overline{\bm{R}}_t
                    \right\}
                \Big| X_t = s_t, \overline{\bm{R}}_t
            \right]
            \nonumber \\
        &=
            E^{(m)}_{\bm{W}^{ex}_t} \left[
                E_{\calE_t} \left\{
                    E_{\overline{\bm{X}}_t} \left(
                        Y^{s_t \overline{\bm{X}}_t}_t
                        \Big| \calE_t, \bm{W}^{ex}_t, X_t = s_t, \overline{\bm{R}}_t
                    \right)
                \right\}
                \Big| X_t = s_t, \overline{\bm{R}}_t
            \right]
            \text{ because } \calE_t \text{ is completely random}
            \nonumber \\
        &=
            E^{(m)}_{\bm{W}^{ex}_t} \left[
                E_{\calE_t} \left\{
                    E_{\overline{\bm{X}}_t} \left(
                        Y^{s_t \overline{\bm{X}}_t}_t
                        \Big| \calE_t, \bm{W}^{ex}_t, \overline{\bm{R}}_t = \bm{1}
                    \right)
                \right\}
                \Big| X_t = s_t, \overline{\bm{R}}_t = \bm{1}
            \right]
            \text{ when } \overline{\bm{R}}_t = \bm{1}
            \nonumber \\
        &=
            E^{(m)}_{\bm{W}^{ex}_t} \left[
                E_{\calE_t} \left\{
                    E_{\overline{\bm{X}}_t} \left(
                        Y^{s_t \overline{\bm{X}}_t}_t
                        \Big| \calE_t, \bm{W}^{ex}_t, \overline{\bm{R}}_t = \bm{1}
                        \right)
                    \right\}
                \Big| \overline{\bm{R}}_t = \bm{1}
            \right]
            \text{ because } \bm{W}^{ex}_t \indep X_t
            \nonumber \\
        &=
            E^{(m)}_{\bm{W}^{ex}_t} \left\{
                E_{\calE_t} \left(
                    Y^{s_t \bigcdot}_t
                \right)
                \big| \overline{\bm{R}}_t = \bm{1}
            \right\}
            \text{ by \eqref{eqn:capo_explicit}; note that } Y^{s_t \bigcdot}_t
                    \text{ still depends on } \overline{\bm{R}}_t
            \nonumber \\
        &=
            E^{(m)}_{\bm{W}^{ex}_t} \left[
                E_{\calE_t} \left\{
                    g_{s_t \bigcdot} \left( \bm{W}^{ex}_t, \calE_t, \overline{\bm{R}}_t = \bm{1} \right)
                \right\}
                \big| \overline{\bm{R}}_t = \bm{1}
            \right]
            \text{ by PO mechanism definition}
            \nonumber \\
        &=
            \frac{1}{m} \sum_{t=1}^m
                E_{\calE_t} \left\{
                    g_{s_t \bigcdot} \left( \bm{W}^{ex}_t, \calE_t, \overline{\bm{R}}_t = \bm{1} \right)
                \right\}
            \nonumber \\
        &=
            E \left(
                Y^{s_t \bigcdot}_t
                \big| \overline{\bm{R}}_{m+1} = \bm{1}
            \right)
\end{align*}
}

\noindent
The empirical mean CAPO in terms of the SCM by adapting equation \eqref{eqn:eyxformula} is derived as follows:
\begin{align*}
    & E \left(
        Y_t
        \big| X_t = s_t, \overline{\bm{R}}_t = \bm{1}
    \right)
    \nonumber \\
        &=
            E^{(m)}_{\bm{W}^{en}_t, \calE_t, \bm{W}^{ex}_t} \left(
                Y_t
                \big| X_t = s_t, \overline{\bm{R}}_t = \bm{1}
            \right)
            \nonumber \\
        &=
            E^{(m)}_{\bm{W}^{ex}_t} \left\{
                E_{\bm{W}^{en}_t, \calE_t} \left(
                    Y_t
                    \big| \bm{W}^{ex}_t, X_t = s_t, \overline{\bm{R}}_t = \bm{1}
                \right)
                \big| X_t = s_t, \overline{\bm{R}}_t = \bm{1}
            \right\}
            \nonumber \\
        &=
            E^{(m)}_{\bm{W}^{ex}_t} \left[
                E_{\calE_t} \left\{
                    E_{\bm{W}^{en}_t} \left(
                        Y_t
                        \big| \calE_t, \bm{W}^{ex}_t, X_t = s_t, \overline{\bm{R}}_t = \bm{1}
                        \right)
                    \big| \bm{W}^{ex}_t, X_t = s_t, \overline{\bm{R}}_t = \bm{1}
                \right\}
                \big| X_t = s_t, \overline{\bm{R}}_t = \bm{1}
            \right]
            \nonumber \\
        &=
            E^{(m)}_{\bm{W}^{ex}_t} \left[
                E_{\calE_t} \left\{
                    E_{\bm{W}^{en}_t} \left(
                        Y_t
                        \big| \calE_t, \bm{W}^{ex}_t, X_t = s_t, \overline{\bm{R}}_t = \bm{1}
                    \right)
                \right\}
                \big| X_t = s_t, \overline{\bm{R}}_t = \bm{1}
            \right]
            \text{ because } \calE_t \text{ is completely random}
            \nonumber \\
        &=
            E^{(m)}_{\bm{W}^{ex}_t} \left[
                E_{\calE_t} \left\{
                    E_{\bm{W}^{en}_t} \left(
                        Y_t
                        \big| \calE_t, \bm{W}^{ex}_t, X_t = s_t, \overline{\bm{R}}_t = \bm{1}
                    \right)
                \right\}
                \big| \overline{\bm{R}}_t = \bm{1}
            \right]
            \text{ because } \bm{W}^{ex}_t \indep X_t
            \nonumber \\
        &=
            E^{(m)}_{\bm{W}^{ex}_t} \left(
                E_{\calE_t} \left[
                    E_{\overline{\bm{X}}_t} \left\{
                        g \left( X_t, \overline{\bm{X}}_t, \bm{W}^{ex}_t, \calE_t, \overline{\bm{R}}_t = \bm{1} \right)
                        \big| \calE_t, \bm{W}^{ex}_t, X_t = s_t, \overline{\bm{R}}_t = \bm{1}
                    \right\}
                \right]
                \big| \overline{\bm{R}}_t = \bm{1}
            \right)
            \text{ by SCM definition}
            \nonumber \\
        &=
            E^{(m)}_{\bm{W}^{ex}_t} \left(
                E_{\calE_t} \left[
                    E_{\overline{\bm{X}}_t} \left\{
                        g \left( s_t, \overline{\bm{X}}_t, \bm{W}^{ex}_t, \calE_t, \overline{\bm{R}}_t = \bm{1} \right)
                        \big| \overline{\bm{R}}_t = \bm{1}
                    \right\}
                \right]
                \big| \overline{\bm{R}}_t = \bm{1}
            \right)
            \nonumber \\
        &=
            E \left(
                Y^{s_t \bigcdot}_t
                \big| \overline{\bm{R}}_{m+1} = \bm{1}
            \right)
            \text{ by \eqref{eqn:eytbarxt_nof1}}
\end{align*}

\noindent
The explicit APTE is derived as follows:
\begin{align}
    \delta^\text{APTE}_{(m)}
        &=
            E^{(m)} \left(
                \delta^\text{PTE}_t
                \big| \overline{\bm{R}}_t = \bm{1}
            \right)
            \nonumber \\
        &=
            E^{(m)}_{\overline{\bm{X}}_t, \calE_t, \bm{W}^{ex}_t} \left\{
                \delta^\text{PTE}_t \left( \overline{\bm{X}}_t, \bm{W}^{ex}_t, \calE_t, \overline{\bm{R}}_t = \bm{1} \right)
                \big| \overline{\bm{R}}_t = \bm{1}
            \right\}
            \nonumber \\
        &=
            E^{(m)}_{\bm{W}^{ex}_t} \left[
                E_{\overline{\bm{X}}_t, \calE_t} \left\{
                    \delta^\text{PTE}_t \left( \overline{\bm{X}}_t, \bm{W}^{ex}_t, \calE_t, \overline{\bm{R}}_t = \bm{1} \right)
                    \big| \bm{W}^{ex}_t, \overline{\bm{R}}_t = \bm{1}
                \right\}
                \big| \overline{\bm{R}}_t = \bm{1}
            \right]
            \nonumber \\
        &=
            \frac{1}{m} \sum_{t=1}^m
            E_{\overline{\bm{X}}_t, \calE_t} \left\{
                \delta^\text{PTE}_t \left( \overline{\bm{X}}_t, \bm{W}^{ex}_t, \calE_t, \overline{\bm{R}}_t = \bm{1} \right)
                \big| \bm{W}^{ex}_t, \overline{\bm{R}}_t = \bm{1}
            \right\}
            \nonumber \\
        &=
            \frac{1}{m} \sum_{t=1}^m
            E_{\overline{\bm{X}}_t, \calE_t} \left\{
                \delta_t \left( \overline{\bm{X}}_t, \bm{W}^{ex}_t, \calE_t, \overline{\bm{R}}_t = \bm{1} \right)
                \big| \bm{W}^{ex}_t, \overline{\bm{R}}_t = \bm{1}
            \right\}
            \text{ by historical PTE definition}
            \nonumber \\
        &=
            \frac{1}{m} \sum_{t=1}^m
            E_{\calE_t} \left[
                E_{\overline{\bm{X}}_t} \left\{
                    \delta_t \left( \overline{\bm{X}}_t, \bm{W}^{ex}_t, \calE_t, \overline{\bm{R}}_t = \bm{1} \right)
                    \big| \calE_t, \bm{W}^{ex}_t, \overline{\bm{R}}_t = \bm{1}
                \right\}
                \big| \bm{W}^{ex}_t, \overline{\bm{R}}_t = \bm{1}
            \right]
            \nonumber \\
        &=
            \frac{1}{m} \sum_{t=1}^m
            E_{\calE_t} \left[
                E_{\overline{\bm{X}}_t} \left\{
                    \delta_t \left( \overline{\bm{X}}_t, \bm{W}^{ex}_t, \calE_t, \overline{\bm{R}}_t = \bm{1} \right)
                    \big| \overline{\bm{R}}_t = \bm{1}
                \right\}
                \big| \bm{W}^{ex}_t, \overline{\bm{R}}_t = \bm{1}
            \right]
            \nonumber \\
        &=
            \frac{1}{m} \sum_{t=1}^m
            E_{\calE_t} \left[
                E_{\overline{\bm{X}}_t} \left\{
                    \delta_t \left( \overline{\bm{X}}_t, \bm{W}^{ex}_t, \calE_t, \overline{\bm{R}}_t = \bm{1} \right)
                    \big| \overline{\bm{R}}_t = \bm{1}
                \right\}
            \right]
            \text{ because } \calE_t \text{ is completely random}
            \nonumber \\
        &=
            \frac{1}{m} \sum_{t=1}^m
            E_{\calE_t} \left\{
                \delta_t \left( \bigcdot, \bm{W}^{ex}_t, \calE_t, \overline{\bm{R}}_t = \bm{1} \right)
            \right\}
            \nonumber \\
        &=
            E \left\{
                \delta_t ( \bigcdot )
                \big| \overline{\bm{R}}_{m+1} = \bm{1}
            \right\}
            \nonumber \\
        &=
            \delta^\text{HAPTE}_{(m)}
        \label{eqn:apte_explicit_appendix}
\end{align}

	\subsubsection*{A3.2.3 Order-1 Model with Randomized Treatments}

We now derive three long-run averages implied by an ARCO model of lag order $1$ (here, with $\overline{\bm{X}}_t^{\ell^X} = X_{t-1}$ and $\overline{\bm{Y}}_t^{\ell^Y} = Y_{t-1}$) when $X$ is randomized at every period as in an n-of-1 experiment. These averages will allow us to identify and estimate the APTE directly using this simple model's parameters, which we will use in a brief simulation study in Section \ref{sec:sims}.

This order-1 model is $Y_t = Y^{X_t X_{t-1} \overline{\bm{X}}_{t-1} }_t = \beta_0 + \beta_X X_t + \beta_{co} X_{t-1} + \beta_{Xco} X_t X_{t-1} + \beta_{ar} Y_{t-1} + \beta_{Xar} X_t Y_{t-1} + \bm{W}^{ex}_t \bm{\beta}_{ex} + \calE_t$. This is because of the implicit recursive dependence of $Y_t$ on the treatment history beyond $t-1$; i.e., $Y_{t-1}$ is a function of $X_{t-2}$ and $Y_{t-2}$, $Y_{t-2}$ is a function of $X_{t-3}$ and $Y_{t-3}$, etc.

In general, we have $E( Y_t | X_t = x_t ) = \beta_0 + \beta_X x_t + \beta_{co} E( X_{t-1} | X_t = x_t ) + \beta_{Xco} x_t E( X_{t-1} | X_t = x_t ) + \beta_{ar} E( Y_{t-1} | X_t = x_t ) + \beta_{Xar} x_t E( Y_{t-1} | X_t = x_t ) + E( \bm{W}^{ex}_t | X_t = x_t ) \bm{\beta}_{ex}$. Now suppose $X$ is always randomized with probability $\Pr( X = 1 ) = \pi$ such that $E( X_{t-1} | X_t = x_t ) = E( X_{t-1} ) = \Pr( X_{t-1} = 1 ) = \pi$, $E( Y_{t-1} | X_t = x_t ) = E( Y_{t-1} )$, and $E( \bm{W}^{ex}_t | X_t = x_t ) = E( \bm{W}^{ex}_t )$. Furthermore, $E( Y_{t-1} ) = E( Y ) = \mu_Y$ because $\big\{ ( Y_t ) \big\}$ is WSS; likewise, $E( \bm{W}^{ex}_t ) = \bm{\mu}_{ex}$.

Finally, recall from equation \eqref{eqn:eytbarxt} that $E( Y_t | X_t = x_t ) = Y^{x_t \bigcdot}_t$. Hence, long-run effect constancy holds because $\delta^\text{PTE}_t = Y^{1 \bigcdot}_t - Y^{0 \bigcdot}_t = E( Y_t | X_t = 1 ) - E( Y_t | X_t = 0 ) = \beta_X + \beta_{Xco} \pi + \beta_{Xar} \mu_Y$ is constant across all periods in the long run. By \eqref{eqn:apte_appendix}, we have:
\begin{equation*}
\delta^\text{APTE} = \beta_X + \beta_{Xco} \pi + \beta_{Xar} \mu_Y    
\end{equation*}
After re-arranging terms, we also have $E( Y_t | X_t = x_t ) = \gamma + x_t \delta^\text{APTE}$ where $\gamma = \beta_0 + \beta_{co} \pi + \beta_{ar} \mu_Y + \bm{\mu}_{ex} \bm{\beta}_{ex}$.

The long-run mean outcome for the ARCO model of order 1 mentioned in the main text is derived as follows.

\begin{align*}
    \mu_Y
        &= E( Y_t ) \\
        &= E( Y_t | X_t = 1 ) \pi + E( Y_t | X_t = 0 ) ( 1 - \pi ) \\
        &= ( \gamma + \delta^\text{APTE} ) \pi + \gamma ( 1 - \pi ) \\
        &= \gamma \pi + \delta^\text{APTE} \pi + \gamma - \gamma \pi \\
        &= \delta^\text{APTE} \pi + \gamma \\
        &= \left( \beta_X + \beta_{Xco} \pi + \beta_{Xar} \mu_Y \right) \pi + \beta_0 + \beta_{co} \pi + \beta_{ar} \mu_Y + \bm{\mu}_{ex} \bm{\beta}_{ex} \\
        &= \beta_X \pi + \beta_{Xco} \pi^2 + \beta_{Xar} \mu_Y \pi + \beta_0 + \beta_{co} \pi + \beta_{ar} \mu_Y + \bm{\mu}_{ex} \bm{\beta}_{ex} \\
    \mu_Y - \beta_{Xar} \mu_Y \pi - \beta_{ar} \mu_Y
        &= \beta_X \pi + \beta_{Xco} \pi^2 + \beta_0 + \beta_{co} \pi + \bm{\mu}_{ex} \bm{\beta}_{ex} \\
    \mu_Y - \beta_{ar} \mu_Y - \beta_{Xar} \mu_Y \pi
        &= \beta_0 + \beta_X \pi + \beta_{co} \pi + \beta_{Xco} \pi^2 + \bm{\mu}_{ex} \bm{\beta}_{ex} \\
    \mu_Y \left( 1 - \beta_{ar} - \beta_{Xar} \pi \right)
        &= \beta_0 + \beta_X \pi + \beta_{co} \pi + \beta_{Xco} \pi^2 + \bm{\mu}_{ex} \bm{\beta}_{ex} \\
    \mu_Y
        &= \frac{\beta_0 + \beta_X \pi + \beta_{co} \pi + \beta_{Xco} \pi^2 + \bm{\mu}_{ex} \bm{\beta}_{ex}}{1 - \beta_{ar} - \beta_{Xar} \pi}
\end{align*}

\section*{A4 Model-Twin Randomization}

	\subsection*{A4.1 MoTR Procedure}

The algebraic identities \eqref{eqn:identity_empirical_mean} and \eqref{eqn:identity_empirical_sd} below can be used to speed up MoTR. A number of summary statistics must be calculatec by MoTR, each over a vector of runs that gets larger with every run. Instead, the following identities can be used to calculate each summary statistic using only two elements: the previously summary statistic (over all completed runs), and the value at the current run of the variable being summarized.

Let $\bar{U}_n$ denote the empirical mean of a variable $U$ over $i = 1, ..., n$ independently identically distributed random samples. We have:

\begin{align*}
    \bar{U}_1
        &=      \frac{1}{1} \left( U_1 \right) \\
    \bar{U}_2
        &=      \frac{1}{2} \left( U_1 + U_2 \right)
        &&=     \frac{1}{2} U_1 +
                \frac{1}{2} U_2 \\
        &=      \left( \frac{1}{1} \cdot \frac{1}{1} \right) \frac{1}{2} \left( U_1 \right) +
                \frac{1}{2} U_2
        &&=     \frac{1}{2} \left\{ \frac{1}{1} \left( U_1 \right) \right\} +
                \frac{1}{2} U_2
        &&&=    \frac{1}{2} \bar{U}_1 +
                \frac{1}{2} U_2
        &&&&=   \frac{1}{2} \left( 1 \bar{U}_1 + U_2 \right) \\
    \bar{U}_3
        &=      \frac{1}{3} \left( U_1 + U_2 + U_3 \right)
        &&=     \frac{1}{3} \left( U_1 + U_2 \right) +
                \frac{1}{3} U_3 \\
        &=      \left( \frac{1}{2} \cdot \frac{2}{1} \right) \frac{1}{3} \left( U_1 + U_2 \right) +
                \frac{1}{3} U_3
        &&=     \frac{2}{3} \left\{ \frac{1}{2} \left( U_1 + U_2 \right) \right\} +
                \frac{1}{3} U_3 
        &&&=    \frac{2}{3} \bar{U}_2 +
                \frac{1}{3} U_3
        &&&&=   \frac{1}{3} \left( 2 \bar{U}_2 + U_3 \right) \\
        &       \vdots
\end{align*}
\begin{equation}\label{eqn:identity_empirical_mean}
    \bar{U}_n
        = I \left( n = 1 \right) U_n + I \left( n > 1 \right) \frac{1}{n} \left\{ (n-1) \bar{U}_{n-1} + U_n \right\}
\end{equation}

\noindent
This derivation was verified by ChatGPT-4o on 2024-06-19 (chat record: \href{https://chatgpt.com/share/32ce253c-40c9-4f8d-9a91-1f2a4c64a829}{https://chatgpt.com/share/32ce253c-40c9-4f8d-9a91-1f2a4c64a829}; copy available in relevant eponymous supplementary file).

Let $\hat{\sigma}_n$ denote the empirical standard deviation of a variable $U$ over $i = 1, ..., n$ independently identically distributed random samples. For $n>2$, we have the following derivation largely provided by ChatGPT-4o between 2024-06-19 and 2024-06-21 (chat record: \href{https://chatgpt.com/share/2484ce8a-6b4d-4e06-8aa4-f73d924abe8c}{https://chatgpt.com/share/2484ce8a-6b4d-4e06-8aa4-f73d924abe8c}; copy available in relevant eponymous supplementary file). This derived formula was verified as correct (i.e., with only minuscule floating-point error) against the \lq\lq ground truth" values calculated using the empirical standard deviation formula.

\bigskip
\noindent
Calculate \(\bar{U}_n\) in terms of \(\bar{U}_{n-1}\):

\[
\bar{U}_n = \frac{1}{n} \sum_{i=1}^n U_i = \frac{1}{n} \left( \sum_{i=1}^{n-1} U_i + U_n \right) = \frac{1}{n} \left( (n-1)\bar{U}_{n-1} + U_n \right)
\]
\[
\bar{U}_n = \frac{n-1}{n} \bar{U}_{n-1} + \frac{1}{n} U_n
\]

\noindent
Express the sum of squared deviations for \(n\) observations. We need to express \(\sum_{i=1}^n ( U_i - \bar{U}_n )^2\) in terms of \(\sum_{i=1}^{n-1} ( U_i - \bar{U}_{n-1} )^2\):

\[
\sum_{i=1}^n ( U_i - \bar{U}_n )^2 = \sum_{i=1}^{n-1} ( U_i - \bar{U}_n )^2 + ( U_n - \bar{U}_n )^2
\]

\noindent
Expand the first term:

\[
\sum_{i=1}^{n-1} ( U_i - \bar{U}_n )^2 = \sum_{i=1}^{n-1} \left( U_i - \bar{U}_{n-1} + \bar{U}_{n-1} - \bar{U}_n \right)^2
\]
\[
= \sum_{i=1}^{n-1} \left[ ( U_i - \bar{U}_{n-1} )^2 + 2 ( U_i - \bar{U}_{n-1} )( \bar{U}_{n-1} - \bar{U}_n ) + ( \bar{U}_{n-1} - \bar{U}_n )^2 \right]
\]

\noindent
Since \(\sum_{i=1}^{n-1} ( U_i - \bar{U}_{n-1} ) = 0\):
\[
\sum_{i=1}^{n-1} ( U_i - \bar{U}_n )^2 = \sum_{i=1}^{n-1} ( U_i - \bar{U}_{n-1} )^2 + (n-1) ( \bar{U}_{n-1} - \bar{U}_n )^2
\]

\noindent
Calculate \((\bar{U}_{n-1} - \bar{U}_n)^2\):

\[
\bar{U}_{n-1} - \bar{U}_n = \bar{U}_{n-1} - \left( \frac{n-1}{n} \bar{U}_{n-1} + \frac{1}{n} U_n \right)
\]
\[
= \bar{U}_{n-1} - \frac{n-1}{n} \bar{U}_{n-1} - \frac{1}{n} U_n
\]
\[
= \frac{1}{n} \bar{U}_{n-1} - \frac{1}{n} U_n = \frac{1}{n} (\bar{U}_{n-1} - U_n)
\]

\noindent
Then,
\[
(\bar{U}_{n-1} - \bar{U}_n)^2 = \left( \frac{1}{n} (\bar{U}_{n-1} - U_n) \right)^2 = \frac{1}{n^2} (\bar{U}_{n-1} - U_n)^2
\]

\noindent
Combine the results:

\[
\sum_{i=1}^n ( U_i - \bar{U}_n )^2 = \sum_{i=1}^{n-1} ( U_i - \bar{U}_{n-1} )^2 + (n-1) \frac{1}{n^2} (\bar{U}_{n-1} - U_n)^2 + ( U_n - \bar{U}_n )^2
\]

\noindent
Simplify \((U_n - \bar{U}_n)^2\):

\[
U_n - \bar{U}_n = U_n - \left( \frac{n-1}{n} \bar{U}_{n-1} + \frac{1}{n} U_n \right)
\]
\[
= U_n - \frac{n-1}{n} \bar{U}_{n-1} - \frac{1}{n} U_n = \frac{n}{n} U_n - \frac{n-1}{n} \bar{U}_{n-1} - \frac{1}{n} U_n
\]
\[
= \frac{n-1}{n} (U_n - \bar{U}_{n-1})
\]

\noindent
Then,
\[
(U_n - \bar{U}_n)^2 = \left( \frac{n-1}{n} (U_n - \bar{U}_{n-1}) \right)^2 = \frac{(n-1)^2}{n^2} (U_n - \bar{U}_{n-1})^2
\]

\noindent
Combine everything to find \(\hat{\sigma}_n\):

\[
\sum_{i=1}^n ( U_i - \bar{U}_n )^2 = \sum_{i=1}^{n-1} ( U_i - \bar{U}_{n-1} )^2 + \frac{n-1}{n^2} (\bar{U}_{n-1} - U_n)^2 + \frac{(n-1)^2}{n^2} (U_n - \bar{U}_{n-1})^2
\]

\noindent
Notice that the last two terms can be combined:

\[
\frac{n-1}{n^2} (\bar{U}_{n-1} - U_n)^2 + \frac{(n-1)^2}{n^2} (U_n - \bar{U}_{n-1})^2 = \frac{(n-1)}{n^2} ( \bar{U}_{n-1} - U_n )^2 \left( 1 + (n-1) \right)
\]

\[
= \frac{(n-1)}{n^2} ( \bar{U}_{n-1} - U_n )^2 ( n ) = \frac{(n-1)}{n} ( \bar{U}_{n-1} - U_n )^2
\]

Thus, the formula for \(\hat{\sigma}_n^2\) in terms of \(U_n\) and \(\hat{\sigma}_{n-1}^2\) is:

\[
\hat{\sigma}_n^2 = \frac{1}{n-1} \left[ \sum_{i=1}^{n-1} ( U_i - \bar{U}_{n-1} )^2 + \frac{n-1}{n} ( U_n - \bar{U}_{n-1} )^2 \right]
\]

\[
\hat{\sigma}_n^2 = \frac{1}{n-1} \left[ (n-2)\hat{\sigma}_{n-1}^2 + \frac{n-1}{n} ( U_n - \bar{U}_{n-1} )^2 \right]
\]

\noindent
Therefore, 

\begin{equation}\label{eqn:identity_empirical_sd}
\hat{\sigma}_n = \sqrt{ \frac{1}{n-1} \left[ (n-2)\hat{\sigma}_{n-1}^2 + \frac{n-1}{n} ( U_n - \bar{U}_{n-1} )^2 \right] }
\end{equation}

\section*{A5 Simulation Study}

	\subsection*{A5.1 Data-generating Procedure}

            \subsubsection*{A5.1.1 Logit-Normal Random Variable}

We will show that $\pi_{ht}$ is a logit-normal random variable for $t>1$, where $\pi_{ht} = \Pr \big( X_{ht} = 1 | Y_{h(t-1)} \big)$. Suppress the $h$ index for clarity.

Recall that $Y^s_t = \beta_0 + \beta_X s + \beta_{ar} Y_{t-1} + \calE_t$ where $\calE_t \sim N \big( 0, \sigma_\varepsilon \big)$. Hence for fixed $X_{t-1} = x_{t-1}$ (by causal consistency) and fixed $Y_{t-2} = y_{t-2}$, we have $Y_{t-1} = \beta_0 + \beta_X x_{t-1} + \beta_{ar} y_{t-2} + \calE_{t-1}$ and therefore $Y_{t-1} \sim N ( \beta_0 + \beta_X x_{t-1} + \beta_{ar} y_{t-2}, \sigma_\varepsilon )$. Because $\pi_t = \text{logistic} \big( Y_{t-1} \big)$, $\pi_t$ is logit-normally distributed.

	\subsection*{A5.2 Analysis Methods and Results}

The stopping rule we followed relies on the coefficient of variation, modified to use the standard error (SE) rather than standard deviation as follows. For $s \in \{ 0, 1 \}$ at MoTR run $r$, calculate the empirical standard deviation of the PO as the standard deviation of the noisy predicted outcomes, denoted $\sigma^s_r$. Calculate the standard error of the estimated mean PO as $SE^s_r = \frac{\sigma^s_r}{\sqrt{m_s}}$. Finally, calculate the SE-based coefficient of variation of the estimated mean PO as $SECV^s_r = \frac{SE^s_r}{ | \bar{\hat{Y}}^s_r | }$.

At MoTR run $r$, if both $SECV^0_r$ and $SECV^1_r$ are less than or equal to some set stopping value $SECV_{stop}$, stop running MoTR and set $r$ as the final run. Report the cumulative values calculated at that run as the APTE estimate $\hat{\delta}^\text{MoTR}$ along with its CI. In both our Simulation Study and Empirical Study, we set $SECV_{stop} = 0.01$.

\setlength{\extrarowheight}{4pt}
\begin{table}
\caption{ \label{tab:sims_results_all_2922} Simulation study analysis results for 100 simulated datasets ($m = 2922$ each) with a true APTE of 1.89.}
\begin{center}
\scalebox{1}{
\begin{tabular}{ l | l }
\hline
Method & Estimated Mean Bias (95\% CI) \\
\hline
Raw Comparison & $-0.15 \; (-0.16, -0.14)$ \\
MoTR-GLM & $-0.00 \; (-0.01, \; 0.00)$ \\
MoTR-RF & $-0.08 \; (-0.09, -0.08)$ \\
PSTn-GLM & $-0.01 \; (-0.03, \; 0.02)$ \\
PSTn-RF & $-0.00 \; (-0.03, 0.02)$ \\
\hline
\end{tabular}
}
\end{center}
\end{table}
\setlength{\extrarowheight}{0pt}

\begin{figure} 
    \centering
    \includegraphics[width=1\linewidth]{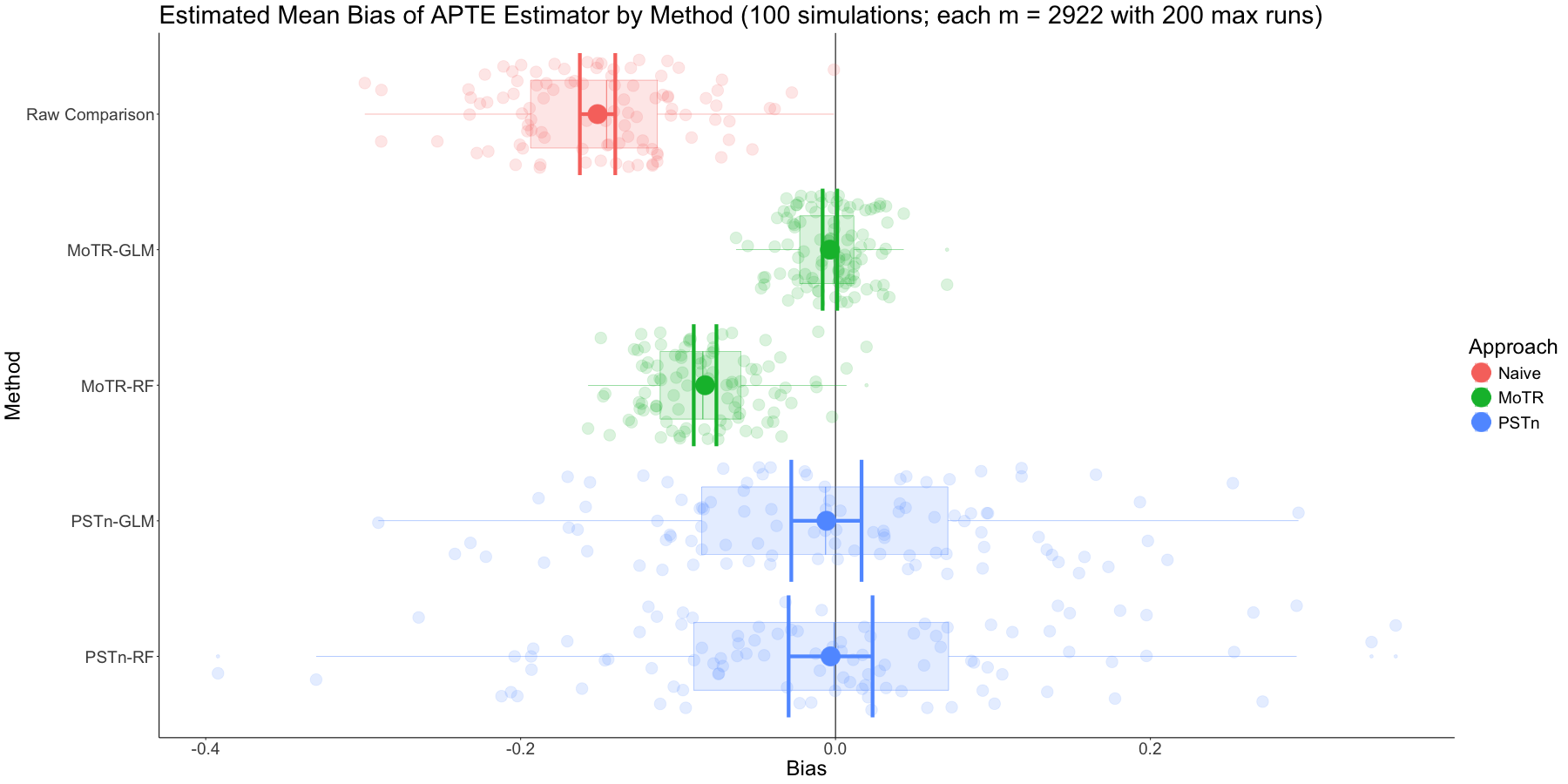}
    \caption{Simulation study analysis results for 100 simulated datasets ($m = 2922$ each) with a true APTE of 1.89. Each small light dot represents the bias of one simulated dataset for estimating a true APTE of 1.89. Each big dark dot represents the average bias (i.e., estimated mean bias) over all 100 datasets, with corresponding 95\% confidence interval shown as symmetric error bars.}
    \label{fig:sims_100_2292}
\end{figure}

\section*{A7 Discussion}

\subsection*{A7.1 Summary of Findings}

See Appendix Table \ref{tab:sims_results_all_2922} and Appendix Figure \ref{fig:sims_100_2292}.

\subsection*{A7.2 Limitations, Extensions, and Future Directions}

\subsubsection*{A7.2.1 Data-Driven Procedure}

\begin{enumerate}

    \item The study participant, their health provider, and the analyst decide on plausible confounders.
    \item The analyst fits and selects or cross-validates initial outcome or propensity models. The analyst then selects the final models; e.g., test all selected models on a holdout set once.
    \item The analyst runs MoTR or PSTn using these final models.
    \item The analyst reports the findings with the largest, most statistically discernible differences (i.e., with the most statistically significant p-values) from the naive estimates. These findings may indicate situations in which confounding is strong enough to change effect estimates---and thereby change how future within-individual interventions are designed.
    \item The analyst discusses the plausibility of the selected models with the study participant and their health provider to determine an intervention plan.
    
\end{enumerate}

\end{document}